\documentclass[letterpaper, 10pt, conference]{ieeeconf} 
\IEEEoverridecommandlockouts
\usepackage{graphicx}
\usepackage{url}
\usepackage{xcolor}
\usepackage{amsmath}
\usepackage{mathtools}
\usepackage{tikz}
\usepackage{verbatim}
\usepackage{subcaption}
\usepackage{epsfig}
\usepackage{pgfplots}
\pgfplotsset{compat=1.10}
\usepackage{graphicx}
\usepackage{epstopdf}
\usepackage{amssymb}
\usepackage{relsize}
\usepackage[textwidth=3.8em,textsize=scriptsize,disable]{todonotes}
\usepackage{algorithm}
\definecolor{Gray}{gray}{0.90}
\usepackage{colortbl}

 \usepackage{url}
 \usepackage{hyperref}
\usepackage{algorithm}
\usepackage{algorithmicx}
\usepackage{algcompatible}
\usepackage{algpseudocode}
\usepackage{color}
\usepackage{blkarray}
\usepackage{setspace} 
\newtheorem{theorem}{\bf{Theorem}}[section]

\newtheorem{lem}[theorem]{\bf{Lemma}}

\newtheorem{remark}[theorem]{\bf{Remark}}
\newenvironment{definition}[1][Definition]{\begin{trivlist}
\item[\hskip \labelsep {\bfseries #1}]}{\end{trivlist}}
\def\qed{\hfill\rule[-1pt]{5pt}{5pt}\par\medskip}
\usepackage{amssymb}
\usepackage{url}
\usepackage{cite}
\begin{document}
\title{Distributed Design of Controllable and Robust Networks using\\ Zero Forcing and Graph Grammars}
\author{Priyanshkumar I. Patel$^\ast$, Johir Suresh$^\ast$, and Waseem Abbas
\thanks{$^\ast$Authors contributed equally to the work.}
\thanks{Authors are with the University of Texas at Dallas, Richardson, TX, USA (Emails: \texttt{priyanshkumar.patel@utdallas.edu, johir.suresh@utdallas.edu, waseem.abbas@utdallas.edu}).}
}

\maketitle
\begin{abstract}
This paper studies the problem of designing networks that are strong structurally controllable, and robust simultaneously. For given network specifications, including the number of nodes $N$, the number of leaders $N_L$, and diameter $D$, where $2 \le D \le N/N_L$, we propose graph constructions generating strong structurally controllable networks. We also compute the number of edges in graphs, which are maximal for improved robustness measured by the algebraic connectivity and Kirchhoff index. For the controllability analysis, we utilize the notion of zero forcing sets in graphs. Additionally, we present graph grammars, which are sets of rules that agents apply in a distributed manner to construct the graphs mentioned above. We also numerically evaluate our methods. This work exploits the trade-off between network controllability and robustness and generates networks satisfying multiple design criteria.
\end{abstract}


\section{Introduction}
\label{sec:intro} 
The distributed design of networks satisfying multiple design criteria is generally a challenging problem. From a network control perspective, controllability and robustness to failures are two of the vital design attributes. Network controllability is the ability to manipulate and drive the network to desired configurations (states) due to external control signals (inputs), which are injected into the network through a subset of nodes called \emph{leaders}~(e.g.,~\cite{pasqualetti2014controllability}). Network robustness has many interpretations, which can be categorized as functional and structural robustness~\cite{abbas2012}. The former is related to the network's functioning in the presence of noise and perturbations and later describes the ability to preserve the network's structural attributes despite node/edge failures~\cite{zelazo2015robustness,young2015new}. Interestingly, these two interpretations are related to each other in the context of network control systems and can be measured through common graph metrics, such as algebraic connectivity and Kirchhoff index $K_f$ \cite{ellens2011effective,wang2010graphs,siami2013fundamental}. 

It is well studied that network controllability and robustness can be conflicting, i.e., for a given set of network parameters, networks requiring few leaders for complete controllability might exhibit poor robustness properties~\cite{abbas2020tradeoff,pasqualetti2020fragility}. For instance, for a given number of nodes $N$, path graphs require a single leader node for complete controllability; however, they have minimum robustness. Similarly, fixing $N$ and diameter $D$, networks with maximum robustness (as measured by the algebraic connectivity and $K_f$) are clique chains~\cite{wang2010graphs,ellens2011effective}; however, they require many leaders $(N-D)$ for complete controllability~\cite{abbas2020tradeoff}. So, an important issue is, \emph{how can we design networks in a distributed manner such that networks can be controlled with few leaders (inputs) and exhibit high robustness simultaneously?} This question becomes more intriguing when the network controllability is considered in the strong structural sense due to computational complexity issues~(e.g., \cite{chapman2013strong,shabbir2022computation,menara2017number,jia2019strong}). Network controllability generally depends on the edge weights; however, edge weights are inconsequential in the case of strong structural controllability (SSC), which essentially depends on the network structure and the leader set. 

In this paper, we propose distributed designs of networks that are strong structurally controllable for a given number of nodes $N$ and leaders $N_L$. At the same time, these networks are robust due to maximal edge sets. For distributed construction of such networks, we utilize graph grammars~\cite{klavins2007programmable,yim2007modular,abbas2011hierarchical}. To ensure SSC, we use the relationship between the notion of zero forcing in graphs and SSC~\cite{monshizadeh2014zero,mousavi2017structural}. Our proposed designs are flexible in the sense that for fixed $N$ and $N_L$, they can produce graphs with varying graph parameters such as the diameter $D$ and robustness, as measured by the Kirchhoff index and algebraic connectivity of graphs while ensuring that graphs remain strong structurally controllable. Thus, the network constructions exploit the trade-off between network controllability and robustness. Our main contributions are summarized below:

\begin{itemize}
    \item For given $N$ (total number of nodes) and $N_L$ (number of leaders), we construct strong structurally controllable graphs with $N_L$ leaders and maximal edge sets. For SSC, we utilize the idea of zero forcing sets. 
    \item Our designs enable generating graphs with diameter $D$, where $2\le D \le N/N_L$, while ensuring that each such graph has a maximal edge set and is strong structurally controllable with $N_L$ leaders. Since network diameter influences its robustness, we can attain networks with various robustness. 
    We also numerically evaluate the robustness of such graphs using algebraic connectivity and Kirchhoff index metrics.
    
    \item Furthermore, we provide distributed ways to construct the above graphs using graph grammars, a set of rules that nodes implement locally to achieve the desired network structure. Finally, we numerically evaluate the proposed schemes. 
\end{itemize}

Our problem setting is similar to the one in \cite{abbas2020tradeoff}, albeit with some significant differences. We use a simpler zero forcing method to analyze strong structural controllability in networks, whereas \cite{abbas2020tradeoff} utilizes graph distances for this purpose. Additionally, for given $N$ and $N_L$, the graphs generated in \cite{abbas2020tradeoff} are of fixed diameter. We provide multiple constructions enabling graphs with different diameters and robustness. Finally, we provide distributed constructions of networks using graph grammars, which are not in \cite{abbas2020tradeoff}.

The rest of the paper is organized as follows: Section~\ref{sec:prelim} presents preliminary ideas and sets up the problem. Section~\ref{subsec:DesNWs} is the main section providing graph constructions for given specifications along with the controllability and robustness analysis of the constructions. Section~\ref{subsec:GGs} provides graph grammars to construct the proposed graphs in a distributed manner. Finally, Section~\ref{sec:conclusion} concludes the paper.

\section{Preliminaries}
\label{sec:prelim}

\subsection{Notations}
\label{subsec:notat}

We consider a \textit{multi-agent network system} as an \textit{undirected graph} $\mathcal{G} = (\mathcal{V},\mathcal{E})$. The \emph{vertex set} $\mathcal{V}$ = $\{v_1,v_2,\dots,v_N\}$ represents the agents (nodes), and the edge set $\mathcal{E} \subseteq \mathcal{V} \times \mathcal{V}$ represents the edges between nodes. We denote the edge between nodes $u$ and $v$ by an unordered pair $(u,v)$. Node $u$ is a \emph{neighbor} of node $v$ if $(u,v)\in\mathcal{E}$. The number of nodes in the neighborhood of $u$ is the \emph{degree} of $u$. The \emph{distance} between nodes $u$ and $v$, denoted by $d(u, v)$, is the number of edges in the shortest path between $u$ and $v$. The \emph{diameter} of $\mathcal{G}$, denoted by $D$, is the maximum distance between any two nodes in $\mathcal{G}$. A \emph{path} of length $k$ is a sequence of nodes that form a subgraph of $\mathcal{G}$, $P_k := <u_0, u_1, u_2, \cdots, u_k >$, where $(u_i,u_{i+1}) \in \mathcal{E},\;\forall i\in\{0,\cdots,k-1\}$. The \emph{leader - follower} system associated with graph $\mathcal{G}$ is defined by the following state-space representation:
\begin{equation} \label{eqn:sysdyn}
    \dot{x}(t) = Mx(t)+Bu(t).
\end{equation}

Here, $x(t) \in \Re^n$ is the state of the system and $M\in\mathcal{M(G)}$ is a system matrix, where $\mathcal{M(G)}$ is a family of 
symmetric matrices associated with an undirected graph $\mathcal{G}$ defined below.
\begin{equation}
    \begin{split}
    \mathcal{M(G)} = \{M \in \Re^{n\times n} &| M = M^T, \text{and for}\:i \neq j,\\ & M_{ij} \neq 0 \Leftrightarrow (i,j) \in \mathcal{E(G)}\}.
\end{split}
\end{equation}
Note that $\mathcal{M(G)}$ includes the adjacency and Laplacian  matrices of $\mathcal{G}$. In \eqref{eqn:sysdyn}, $u(t) \in \Re^m$ is the input signal, and $B \in \Re^{n \times m}$ is the input matrix containing information about the leader nodes through which inputs are injected into the network. For a set of leaders labelled $\{\ell_1,\ell_2,\dots,\ell_m\} \subseteq \mathcal{V}$, we define the input matrix as follows.
\begin{equation}
    B_{ij} = 
    \begin{cases}
      1 & \text{if}\: v_i = \ell_j,\\
      0 & \text{otherwise.}
    \end{cases}
\end{equation}
We are interested in designing networks with the above dynamics that are strong structurally controllable and maximally robust. Next, we discuss the controllability and robustness measures we utilize to evaluate our graphs.
\subsection{Network Controllability Measure}
For the strong structural controllability analysis, we utilize the notion of \emph{zero forcing sets} in graphs. Considering a system defined on graph $\mathcal{G}$, the pair $(M,B)$ is a \emph{controllable pair} if there exists an input $u(t)$ that could drive the system from any initial state $x(t_0)$ to any final state $x(t_f)$ in a given time period $t=t_f-t_0$.

\begin{definition} \emph{(Strong Structural Controllability (SSC))}
A given graph $\mathcal{G}$ with a set of leader nodes $\{\ell_1,\ell_2,\dots,\ell_m\} \subseteq \mathcal{V}$, and the corresponding $B$ matrix is said to be \emph{strong structurally controllable} if and only if $(M,B)$ is a controllable pair $\forall \:M \in \mathcal{M}$.
\end{definition}


In \cite{monshizadeh2014zero}, Monshizadeh et al. provides a graph-theoretic characterization of SSC in networks in terms of zero forcing in graphs explained below.

\begin{definition}\emph{(Zero Forcing Process)}
Consider a graph $\mathcal{G} = (\mathcal{V},\mathcal{E})$ whose nodes are initially colored either black, or white. If a node $v \in \mathcal{V}$ is black and has exactly one white neighbor $u$, then $v$ forces $u$ to change its color to black. Zero forcing is a process of applying this color change rule until no black node exists with only one white neighbor.  
\end{definition}

\begin{figure}[h]
    \centering
    \includegraphics[scale=0.6]{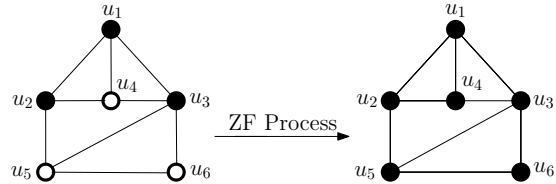}
    \caption{Set of nodes $\{u_1,u_2,u_3\}$ is a ZFS as the corresponding derived set $\mathcal{V}'$ contains all the nodes in the graph.}
    \label{fig:ZFS}
\end{figure}

For a given set of initial black nodes, there can be multiple ways to execute the zero forcing process; however, the set of black nodes at the end of the process will always be the same~\cite{work2008zero}. If there is a unique way of proceeding the zero forcing process in a graph $\mathcal{G}$, we call it a \emph{unique zero forcing process}. Moreover, the set of black nodes obtained at the end of the zero forcing process is called the \emph{derived set}.
\begin{definition}\emph{(Zero Forcing Set (ZFS))}
Consider a graph $\mathcal{G} =(\mathcal{V},\mathcal{E})$ with an initial set of black nodes (leaders) $\{\ell_1,\ell_2,\dots,\ell_m\}\subseteq \mathcal{V}$. Let $\mathcal{V}'$ be the derived set at the of the zero forcing process, then $\{\ell_1,\ell_2,\dots,\ell_m\}$ is a ZFS if and only if $\mathcal{V}'=\mathcal{V}$.

\end{definition}
Figure~\ref{fig:ZFS} illustrates the idea of a ZFS.

Monshizadeh et al.~\cite{monshizadeh2014zero} characterizes the minimum leader set for SSC in terms of ZFS of the network graph, showing that the network is strong structurally controllable if and only if the leader set is a ZFS. In this work, since we aim to design strong structurally controllable networks with $N_L$ leaders, the leader sets will always be zero forcing sets of the corresponding graphs.


\subsection{Network Robustness Measures}
To analyze the robustness of the proposed graphs, we use widely used metrics, \emph{algebraic connectivity} and \emph{Kirchhoff index}. 
\emph{Algebraic connectivity} of a graph $\mathcal{G}$ (also known as the Fiedler value) is the second smallest eigenvalue of its Laplacian matrix. A higher value of algebraic connectivity indicates higher robustness. 

\emph{Kirchhoff index} of a graph $\mathcal{G}$, denoted as $K_f(\mathcal{G})$, is 
\begin{equation}
    K_f(\mathcal{G}) = N \sum_{i=2}^{N} \frac{1}{\lambda_i},
\end{equation}
where $N$ is the total number of nodes in the graph $(\mathcal{G})$ and $\lambda_2 \leq \lambda_3 \leq \cdots \lambda_N$ are the eigenvalues of the Laplacian of the graph $(\mathcal{G})$. Robustness and the value of Kirchhoff index of a graph are inversely related, i.e., lower Kirchhoff index implies higher robustness, and vice versa \cite{young2015new}, \cite{ellens2011effective}, \cite{siami2013fundamental}. We note that network robustness, as measured by both of these measures, is a monotonically increasing function of edge additions~\cite{ellens2011effective,wang2010graphs}. Thus, adding edges to a graph improves its robustness to failures and noise. 
\section{Designing Controllable and Robust Networks}
\label{subsec:DesNWs}
In this section, we will design strong structurally controllable and maximally robust networks for a given number of nodes $N$ and number of leaders $N_L$. We will present three designs, each with different characteristics and performances. In the next section, we provide a distributed way of constructing the proposed graphs; wherein all the nodes follow a set of local rules (\textit{graph grammars}) to make connections with their neighbors to achieve the desired graphs. 


\subsection{Network Design 1}
\label{subsec:CGG1}

\begin{figure}[h]
    \centering
    \includegraphics[scale=0.60]{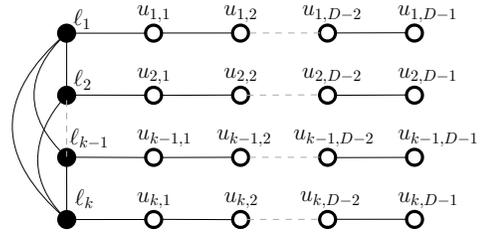}
    \caption{Graph $\mathcal{G}_1$ with $N_L$ number of leaders,  $D$ diameter and $N = (N_L\times D)$ nodes.}
    \label{fig:GG1_Construct_IEEE}
\end{figure}

We know that a  graph with a single leader can only be completely controllable if the graph is a path graph with the leader node being one of the end nodes. Therefore, given $N_L$ number of leaders, and $N = (N_L \times D)$ number of total nodes, where $D$ is the diameter, we construct a graph $\mathcal{G}_1$ with these specifications. For $\mathcal{G}_1$, we create $N_L$ path graphs, each with a diameter $D-1$. Also, the end node of each path is a leader. We then make all leaders pair-wise adjacent, thus, inducing a complete graph among leaders. We describe this construction formally below. Consider the following vertex set for graph $\mathcal{G}_1$:

\begin{center}
    $V = \{\ell_i\} \cup \{u_{i,j}\}$,
\end{center}

where $i \in \{1,2,\dots,k\}$ and $j \in \{1,2,\dots,D-1\}$. Vertices with label $\{\ell_1,\ell_2,\dots,\ell_k\}$ are leaders and the rest $\{u_{1,1},\dots, u_{i,j},\dots, u_{k,D-1}\}$ are followers. Connect the vertices in the following manner:
\begin{itemize}
    \item All the leaders $\ell_i$ have a link between them and generate a complete graph among them.
    \item For all $i \in \{1,2,\dots,k\}$, there exists a link between $\ell_i$ and $u_{i,1}$.
    \item For all $i \in \{1,2,\dots,k\}$ and $j \in \{1,2,\dots,D-2\}$, there is a link between $u_{i,j}$ and $u_{i,j+1}$.
\end{itemize}
Figure~\ref{fig:GG1_Construct_IEEE} illustrates the construction of Graph $\mathcal{G}_1$. 

The graph constructed above is strong structurally controllable with diameter $D-1$. 
Note that we can add edges to the graph $\mathcal{G}_1$ without affecting both, SSC of the graph and the distances between leaders and remaining nodes. Adding edges is useful to increase the graph robustness.

\begin{figure}[h]
    \centering
    \includegraphics[scale=0.60]{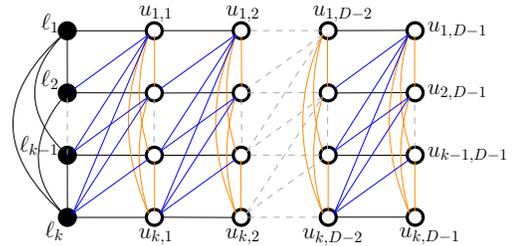}
    \caption{Graph $\bar{\mathcal{G}}_1$ with maximal edges.}
    \label{fig:GG1_Maximal}
\end{figure}

\textit{Adding Maximum Edges to Graph} $\mathcal{G}_1$: Adding edges reduces Kirchhoff index and increases the algebraic connectivity, thus, improving robustness~\cite{ellens2011effective}. We refer the maximally robust graph constructed from $\mathcal{G}_1$ as $\bar{\mathcal{G}}_1$. Note that the addition of edges must not deteriorate SSC. We propose the following addition of edges for the construction of $\bar{\mathcal{G}}_1$:

\begin{itemize}
    \item All the leaders $\ell_i$ has an edge with $u_{q,1}$ $\forall q < i$, where $i,q \in \{1,2,\dots,k\}$.
    \item Similarly, all the nodes in $u_{i,j}$ has an edge with nodes in $u_{q,j+1}$ $\forall q < i$, where $i,q \in \{1,2,\dots,k\}$ and $j \in \{1,2,\dots,D-2\}$.
    \item Also, for a fixed $j$, all nodes in $u_{i,j}$ generate a complete graph, where $i \in \{1,2,\dots,k\}$ and $j \in \{1,2,\dots,D-1\}$
\end{itemize}
Figure~\ref{fig:GG1_Maximal} illustrates the construction of $\bar{\mathcal{G}_1}$ from $\mathcal{G}$. The newly added edges are shown in blue and orange.

\lem \label{lem:ZFSlemmaG1}The leader set $\{\ell_1,\ell_2,\cdots,\ell_{N_L}\}$ is a ZFS of $\bar{\mathcal{G}}_1$ (described above) with $N$ nodes and $D$ diameter.

\proof We observe that in $\bar{\mathcal{G}}_1$, all leaders, except $\ell_1$, have more than one white neighbor in their neighborhoods. So, only $\ell_1$ can initiate the zero forcing process. Leader ($\ell_2$) adjacent to $\ell_1$ has exactly two white neighbors one of which, $u_{1,1}$, it shares with $\ell_1$. Consequently, the third leader has exactly three white neighbors and shares one white neighbor each with $\ell_1$ and $\ell_2$. This continues for all $N_L$ number of leaders.

As discussed, $\ell_1$ starts the zero forcing process by coloring its only white neighbor, $u_{1,1}$. As a result, $\ell_2$ is now left with only one white neighbor, $u_{2,1}$, and thus $\ell_2$ colors it.
Similarly, $\ell_3$ is now left with only one white neighbor that is $u_{3,1}$ because the other two white neighbors $u_{2,1}$ and $u_{1,1}$, which it had in common with $\ell_2$ and $\ell_1$, respectively, are now colored. This continues until all the nodes in $u_{i,1}$ are colored. Note that there exists a complete graph between all the followers in $u_{i,1}$, therefore, $u_{1,1}$ will only be able to color $u_{1,2}$ once all the other nodes in $u_{i,1}$ are colored. We also know that the nodes in $u_{i,1}$ and in $ u_{i,2}$ have similar connections between them as $\ell_i$ and $u_{i,1}$. So, it follows from above discussion that the process continues until all the nodes are colored, implying that the given leader set is a ZFS of $\bar{\mathcal{G}}_1$, which is the desired claim. 
\qed

\lem For fixed number of nodes $N$ and diameter $D$, a graph of construction $\bar{\mathcal{G}}_1$ has maximal edges, i.e., by adding any additional edge the leader set $\{\ell_1,\ell_2,\cdots,\ell_{N_L}\}$ will no longer be a ZFS of $\bar{\mathcal{G}}_1$.

\proof Let us show that the above statement holds for a subgraph $\bar{\mathcal{G}_1'}$ containing only the leader set and the first set of followers $u_{i,1}$ and all the edges between them. As mentioned in Lemma~\ref{lem:ZFSlemmaG1} the zero forcing process in $\bar{\mathcal{G}}_1$ is unique and it propagates from first follower $u_{1,1}$ to all the nodes in $u_{i,1}$ till $u_{k,1}$, in this particular order. Now, considering $\bar{\mathcal{G}_1'}$, we observe that adding any edge would disturb this zero forcing process because an additional edge would result in some leader having more than one white neighbor at a particular time step. This results in the leader set not being a ZFS of the subgraph $\bar{\mathcal{G}_1'}$.
 
 Next, we assume that the above argument is true for all nodes in $\bar{\mathcal{G}}_1$ until the set of nodes in $u_{i,D-2}$. This means that all the nodes in $u_{i,D-2}$ are colored. Since the nodes in $u_{i,D-1}$ are further ahead in the zero forcing process than nodes in $u_{i,D-2}$, we can say that nodes in $u_{i,D-2}$ are not dependent on nodes in $u_{i,D-1}$ for getting colored. Furthermore, we notice that edge set between $u_{i,D-2}$ and $u_{i,D-1}$ is same as that between the leader set and $u_{i,1}$. We use the same reasoning as above to show that we cannot add any other edge between these two node sets without disrupting the zero forcing process, implying that not all the nodes will get colored. Therefore, by induction, adding any extra edge in $\bar{\mathcal{G}}_1$ would result in the leader set not being a ZFS, which is the desired claim. \qed

\remark \label{rem:GPMI} Graph $\bar{\mathcal{G}_1}$ is same (isomorphic) as the graph produced in \cite{abbas2020tradeoff}. We call the graph constructed in \cite{abbas2020tradeoff} as $\mathcal{G}_{PMI}$. It is interesting to note that even though we constructed $\bar{\mathcal{G}_1}$ using the zero forcing method, we arrive at the same result, whereas \cite{abbas2020tradeoff} uses the distance-based approach in their design.

In $\bar{\mathcal{G}}_1$, the diameter is $N/N_L$, i.e., by changing the total number of nodes $N$ and the number of leaders $N_L$, the diameter varies.  So, the interesting question is, \textit{can we design graphs with improved robustness while constraining/fixing the diameter without deteriorating controllability?} In the following subsection, we answer this by designing strong structurally controllable graphs $\bar{\mathcal{G}_2}$ with diameter $D=2$, $N$ total nodes, $N_L\ge 2$ leaders, and improved robustness.


\subsection{Network Design 2}
\label{subsec:CGG2}

We construct a maximally robust graph $\bar{\mathcal{G}_2}$ by fixing $N_L\ge 2$ and adding $N_F = (N - N_L)$ number of other nodes (followers) one-by-one to the graph. This design is different from $\bar{\mathcal{G}_1}$ in that the maximum distance between any two nodes is two. Next, we explain the construction of $\bar{\mathcal{G}}_2$. Consider the following vertex set.

$$V = \{\ell_i\} \cup \{u_j\},$$where $i \in \{1,2,\dots,k\}$ and $j \in \{1,2,\dots,m\}$, where $k = N_L$  and $m = N_F$. Vertices labeled $\{\ell_1, \ell_2, \dots, \ell_k\}$ are leaders and $\{u_1, u_2, \dots, u_m\}$ are followers. 
%
%
%
%

We connect the vertices as follows,
\begin{itemize}
    \item Leader $\ell_1$ and followers $\{u_1, u_2, \dots, u_m\}$ are connected through a path graph starting from $\ell_1$.
    \item All the leaders $\ell_i$ are connected with all the nodes in $u_j$, where $i \in \{2, \dots ,k\}$ and $j \in \{1, \dots ,m\}$
\end{itemize}

\begin{figure}[h]
    \centering
    \includegraphics[scale=0.55]{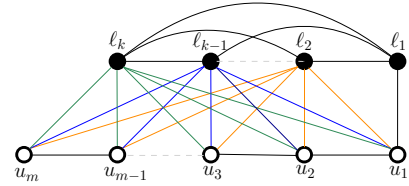}
    \caption{Graph $\bar{\mathcal{G}}_2$ with maximal edges.}
    \label{fig:GG2_maxgen}
\end{figure}

Figure~\ref{fig:GG2_maxgen} illustrates the construction of $\bar{\mathcal{G}}_2$.

\lem \label{lem:zfsg2}For a graph $\bar{\mathcal{G}}_2$ (as described above) with $N$ number of nodes, the proposed leader set $\{\ell_1,\ell_2,\cdots,\ell_{N_L}\}$ is a ZFS.

\proof From the construction of $\bar{\mathcal{G}}_2$, we observe that all leaders, except $\ell_1$, are pair-wise adjacent to all the followers, $u_j$ $\forall j$.
This means that except $\ell_1$, all leaders will generally have more than one white neighbor. Since $\ell_1$ has only one white neighbor $u_1$, it will start the zero forcing process by coloring $u_1$.

Next, the rest of the leaders will still have multiple white neighbors in their neighborhoods. However, $u_1$ has only one white neighbor $u_2$. That will allow $u_1$ to color $u_2$. Subsequently, $u_2$ will also have only one white neighbor $u_3$. This stands true for rest of the follower nodes in $u_j$, where $1 \leq j \leq (N - N_L)$. We note that as a result of this unique zero forcing process, the entire graph gets colored, implying that the leader set is a ZFS of $\bar{\mathcal{G}}_2$. \qed

\lem For fixed number of nodes $N$ and given leader set $\{\ell_1,\ell_2,\cdots,\ell_{N_L}\}$ the graph generated using the construction $\bar{\mathcal{G}}_2$ has maximal edge set, i.e., by adding any additional edge the leader set $\{\ell_1,\ell_2,\cdots,\ell_{N_L}\}$ will no longer be a ZFS of $\bar{\mathcal{G}}_2$.

\proof As shown in Lemma~\ref{lem:zfsg2}, the Graph $\bar{\mathcal{G}}_2$ has a unique zero forcing process. There are two types of edges that we can add to  $\bar{\mathcal{G}}_2$. They include,

\begin{itemize}
    \item leader to follower (non-leader) edges, i.e., ($\ell_1,u_j$), where $j \neq 1$, and
    
    \item follower to follower edges, i.e., ($u_j,u_{j'}$), where $j \neq j'$.
\end{itemize} 
Both categories of edges belong to a unique zero forcing process. Adding an edge between any two non-adjacent nodes in this unique zero forcing process will not allow the preceding node to continue the zero forcing process since it will now have more than one white neighbor. This means that the addition of any edge other than the ones already existing in $\bar{\mathcal{G}}_2$ will result in a graph for which the given leader set is not a ZFS.  \qed

It is interesting to note that, the above two constructions, $\bar{\mathcal{G}}_1$ and $\bar{\mathcal{G}}_2$ generate equal number of edges for the same parameters ($N$ and $N_L$). The number of edges in $\bar{\mathcal{G}}_1$ is,

\begin{equation}
\begin{split}
\label{eq:Edges_G1}
    E_{\bar{\mathcal{G}}_1}  = \underbrace{D\times\frac{N_L\times(N_L-1)}{2}}_{{E}_1} + \underbrace{(D-1)\times\frac{N_L\times(N_L+1)}{2}}_{{E}_2}.
\end{split}
\end{equation}
There are $D$ cliques $\bar{\mathcal{G}}_1$, each of size $N_L$. $E_1$ is the total number of edges in these cliques, and $E_2$ is the number of remaining edges in $\bar{\mathcal{G}}_1$. Similarly, the number of edges in $\bar{\mathcal{G}_2}$ is given by,

\begin{equation}
\begin{split}
\label{eq:Edges_G2}
E_{\bar{\mathcal{G}}_2}  & = \underbrace{(N-N_L)\times\ (N_L-1)}_{{E}_3}+\underbrace{N-N_L}_{{E}_4}\\
& +\underbrace{\frac{N_L\times(N_L-1)}{2}}_{{E}_5}.
\end{split}
\end{equation}

Here, ${E}_3$ is the number of edges between $N_L-1$ leaders and $(N-N_L)$ followers, ${E}_4$ is the number of edges in the path induced by leader $\ell_1$ and followers, and ${E}_5$ is the number of edges in the complete graph induced by the leader nodes. Now, simplifying \eqref{eq:Edges_G1} and \eqref{eq:Edges_G2} gives

\begin{equation}
\label{eq:EdgesG1G2}
E_{\bar{\mathcal{G}}_1}  = E_{\bar{\mathcal{G}}_2} = N_L\times\left(N -\frac{(N_L+1)}{2}\right).
\end{equation}


As discussed previously, the diameter of $\bar{\mathcal{G}}_1$ depends on $N_L$ and $N$, whereas the diameter of $\bar{\mathcal{G}}_2$ is constant regardless of $N_L$ and $N$. So, next, we explore a graph construction that combines $\bar{\mathcal{G}}_1$ and  $\bar{\mathcal{G}}_2$ and affords the option of choosing the diameter $D$ of the graph.
Some applications might require particular diameter values, and by combining $\bar{\mathcal{G}}_1$ and $\bar{\mathcal{G}}_2$, we can have a graph where the diameter $D$ is also a design parameter. We will see in Subsection~\ref{sec:NumEvaRobAnsys} that in many cases $\bar{\mathcal{G}}_2$ provide higher robustness than $\bar{\mathcal{G}}_1$, but require the same amount of leaders to achieve SSC. So, it is advantageous to have $\bar{\mathcal{G}}_1$ combined with $\bar{\mathcal{G}}_2$ for achieving higher robustness while meeting a design requirement in terms of the diameter.

\subsection{Network Design 3 (Combining Designs 1 and 2)}
\label{subsec:CGGmixed}
In this section, we construct a graph that is a combination of $\bar{\mathcal{G}}_1$ and $\bar{\mathcal{G}}_2$, meaning partial nodes follow the rules of construction of $\bar{\mathcal{G}}_1$ and rest of them follow the construction rules of $\bar{\mathcal{G}}_2$ (as discussed in the previous subsections). We define the construction with three different parameters, $N$ (total number of nodes), $N_L$(number of leaders), and $D$ (diameter of the graph). For these given parameters, we construct a graph $\bar{\mathcal{G}_3}$ that is strong structurally controllable, maximally robust and has equal number of edges as $\bar{\mathcal{G}}_1$ or $\bar{\mathcal{G}}_2$ for the same $N$ and $N_L$. 
Let $\bar{\mathcal{V}} = \bar{\mathcal{V}_1} \cup \bar{\mathcal{V}_2}$, be the set of all the nodes in $\bar{\mathcal{G}_3}$, where $\bar{\mathcal{V}_1}$ and $\bar{\mathcal{V}_2}$ are the subset of nodes that follow construction of $\bar{\mathcal{G}}_1$ and $\bar{\mathcal{G}}_2$, respectively. Then, the graph $\bar{\mathcal{G}}_3$ is constructed as follows:
\begin{itemize}
    \item The leader set $\{\ell_1,\ell_2,\cdots,\ell_{N_L}\} \subset \bar{\mathcal{V}}_1$.
    \item Let the end nodes of the construction of $\bar{\mathcal{G}}_1$, i.e., $u_{i,D-2}$, where $1\leq i\leq N_L$, be the pseudo-leaders of $\bar{\mathcal{G}}_2$.
    \item Let $u_{i,D-2} = \bar{\mathcal{V}}_1 \cap \bar{\mathcal{V}}_2 $, where $1\leq i\leq N_L$, then the total number of nodes become $|\bar{\mathcal{V}}| = |\bar{\mathcal{V}_1}|+|\bar{\mathcal{V}_2}|-(\bar{\mathcal{V}}_1 \cap \bar{\mathcal{V}}_2)$. 
    \item The first pseudo-leader of $\bar{\mathcal{G}}_2$, i.e.,$u_{1,D-2}$, belongs to the same zero forcing path as of the first leader ($\ell_1$) in $\bar{\mathcal{G}}_1$.
    \item 
    The edges between nodes $x$ and $y$, where $x\in \{u_{i,D-2},\; \forall i\}$ and $y\in \{u_{i,D-2},\; \forall i\}$, are according to the construction of $\bar{\mathcal{G}}_1$. Similarly, the edges between nodes in $\{u_{i,D-2}, \; \forall i\}$ and $\{v_j\}$, where $1\leq j\leq(|\bar{\mathcal{V}}_2|-N_L)$, are according to the construction of $\bar{\mathcal{G}}_2$.

\end{itemize}



\begin{figure}[htb]
    \centering
    \begin{subfigure}[b]{0.23\textwidth}
	\centering
	\includegraphics[scale=0.5]{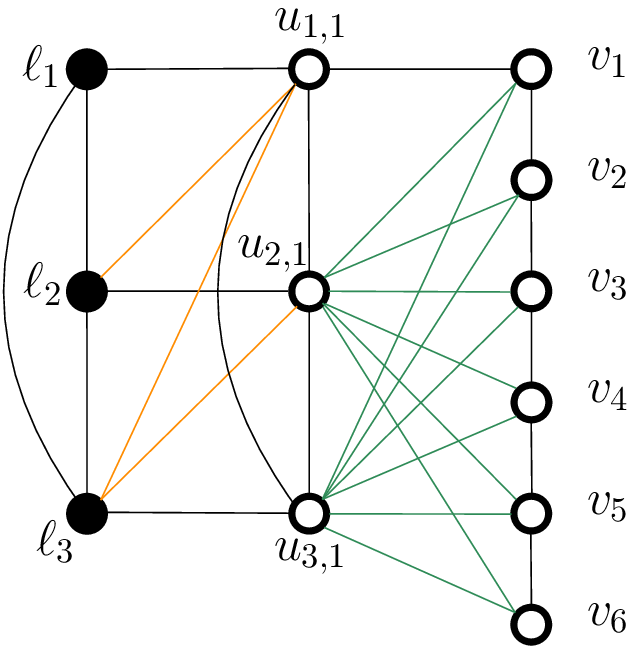}
	\caption{$D=3$}
	\end{subfigure}
	\begin{subfigure}[b]{0.24\textwidth}
	\centering
	\includegraphics[scale=0.5]{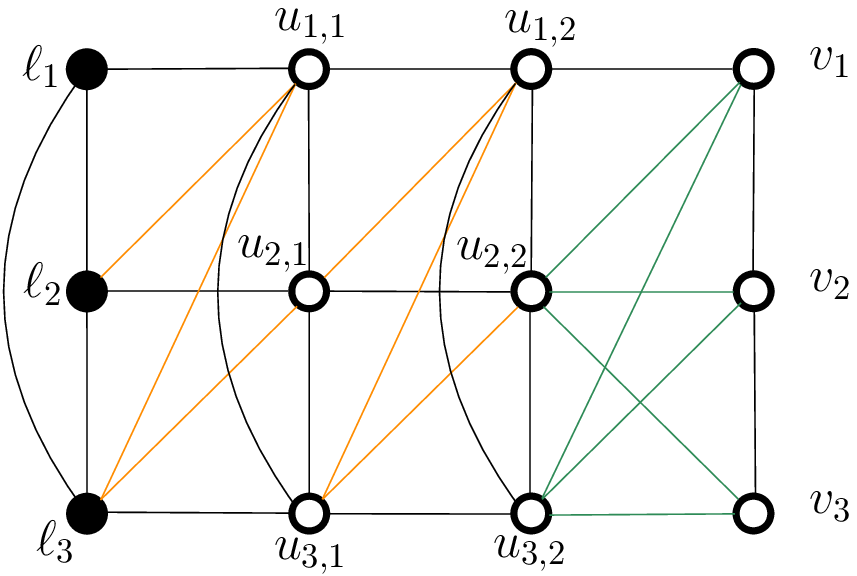}
	\caption{$D=4$}
	\end{subfigure}
    \caption{Examples of network design 3 $(\bar{\mathcal{G}}_3)$.}
    \label{fig:Mixed_graphs}
\end{figure}

Figure~\ref{fig:Mixed_graphs} illustrates two examples of the construction of $\bar{\mathcal{G}_3}$ for $N=12$ and $N_L=3$. The diameters of graphs in (a) and (b) are 3 and 4, respectively.


We make the following observations from the examples:
\begin{itemize}
    \item Both graphs have the same number of edges, which is also equal to the number of edges in graphs generated according to  $\bar{\mathcal{G}}_1$ and  $\bar{\mathcal{G}}_2$ for the same $N$ and $N_L$. 
    \item Both graphs are strong structurally controllable with the given leader sets.
    \item In general, changing $|\bar{\mathcal{V}_1}|$ and $|\bar{\mathcal{V}_2}|$ will result in graphs ranging from diameter $D = 2$ to diameter of $\bar{\mathcal{G}}_1$ for the same $N$ and $N_L$, i.e., 
    \begin{equation}
        2=D(\bar{\mathcal{G}}_2) \le D(\bar{\mathcal{G}}_3) \le D(\bar{\mathcal{G}}_1) = N/N_L.
    \end{equation}
    
    Thus, if $|\bar{\mathcal{V}_1}| = N_L$, $\bar{\mathcal{G}}_3 = \bar{\mathcal{G}}_2$, and similarly, if $|\bar{\mathcal{V}_2}| = N_L$, $\bar{\mathcal{G}}_3 = \bar{\mathcal{G}}_1$.
  
\end{itemize}

\subsection{Numerical Evaluation and Robustness Analysis}
\label{sec:NumEvaRobAnsys}
\begin{figure}[htb]
    \centering
    \begin{subfigure}[b]{0.23\textwidth}
	\centering
	\includegraphics[scale=0.4]{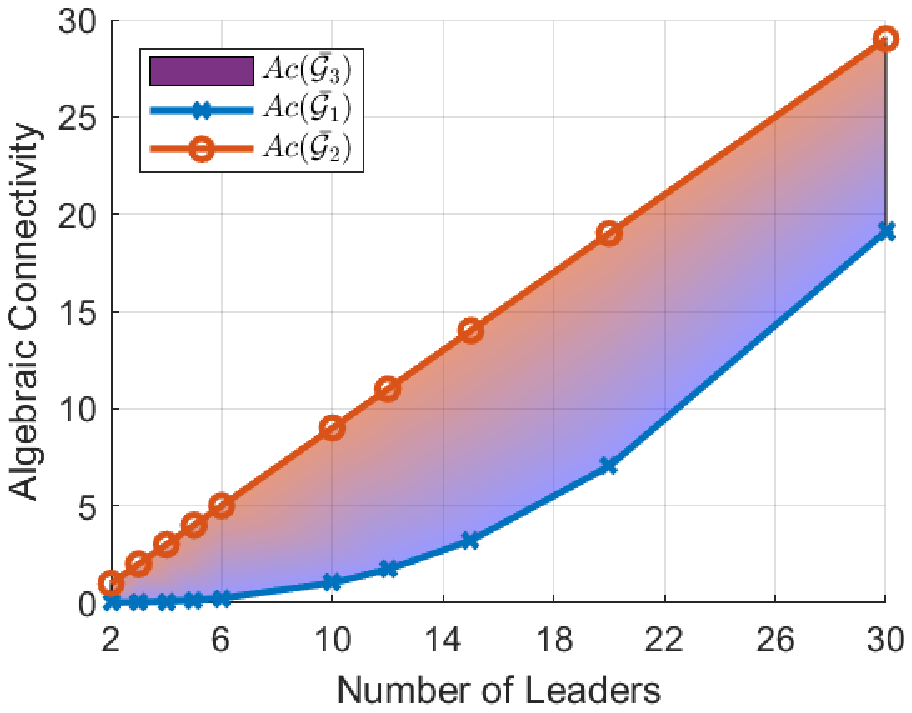}
	\caption{}
	\label{fig:AC_NL_G1_G2_Mix}
	\end{subfigure}
	\begin{subfigure}[b]{0.24\textwidth}
	\centering
	\includegraphics[scale=0.41]{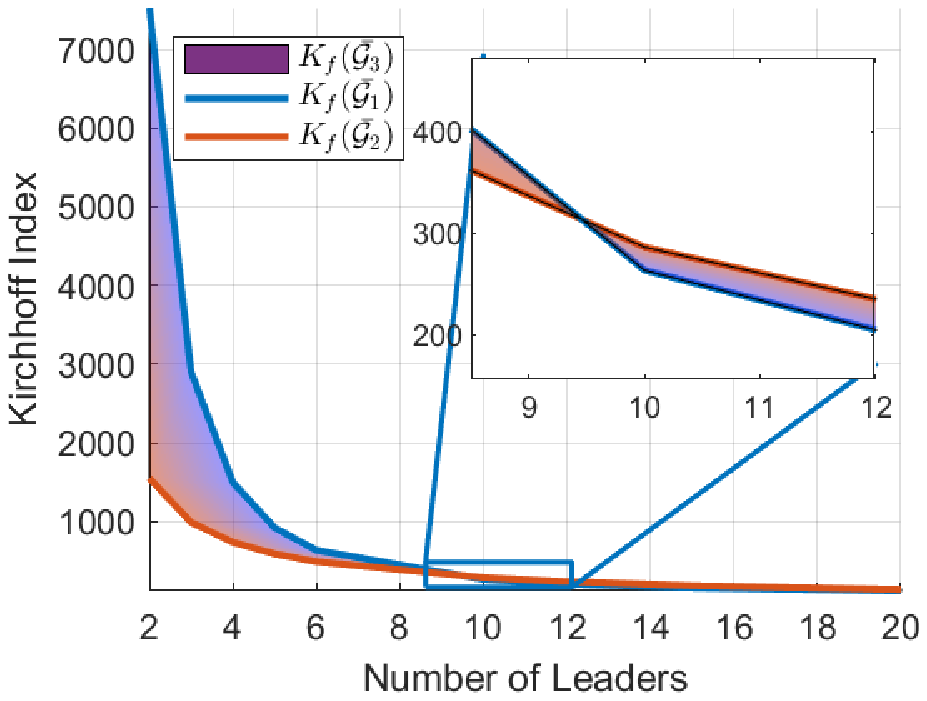}
	\caption{}
	\label{fig:Kf_NL_G1_G2_Mix}
	\end{subfigure}
    \caption{Algebraic connectivity and Kirchhoff index as a function of number of leaders in $\bar{\mathcal{G}}_1$, $\bar{\mathcal{G}}_2$ and $\bar{\mathcal{G}}_3$.}
\end{figure}

Here, we numerically evaluate the performance of graphs $\bar{\mathcal{G}}_1$, $\bar{\mathcal{G}}_2$, $\bar{\mathcal{G}}_3$, in terms of robustness and controllability for $N=60$. First, we analyze the algebraic connectivity of the proposed graphs while varying the number of leaders $N_L$. From Figure~\ref{fig:AC_NL_G1_G2_Mix}, we observe that the algebraic connectivity of $\bar{\mathcal{G}}_2$ is higher than $\bar{\mathcal{G}}_1$ for any given $N_L$. This implies that robustness of  $\bar{\mathcal{G}}_2$, as measured by the algebraic connectivity, is always better than $\bar{\mathcal{G}}_1$. Also, the algebraic connectivity of $\bar{\mathcal{G}}_3$ always lies between $\bar{\mathcal{G}}_2$ and $\bar{\mathcal{G}}_1$.

Figure~\ref{fig:Kf_NL_G1_G2_Mix} plots the robustness performance of the proposed graphs in terms of the Kirchhoff index. Interestingly, for a lower number of leaders, the Kirchhoff index of $\bar{\mathcal{G}}_2$ is significantly lower (indicating improved robustness) than that of $\bar{\mathcal{G}}_1$. However, for a higher number of leaders, this trend changes, and $\bar{\mathcal{G}}_1$ has a lower Kirchhoff index than $\bar{\mathcal{G}}_2$, though the difference between the values remains relatively small. 

Finally, as discussed in the previous subsection, with $\bar{\mathcal{G}}_3$ we can generate graphs whose diameters are lower bounded by $\bar{\mathcal{G}}_2$ and upper bounded by $\bar{\mathcal{G}}_1$. Similarly, the algebraic connectivity and Kirchhoff index for $\bar{\mathcal{G}}_3$ are also bounded by those of $\bar{\mathcal{G}}_1$ and $\bar{\mathcal{G}}_2$. This is shown in Figure~\ref{fig:AC_NL_G1_G2_Mix} and Figure~\ref{fig:Kf_NL_G1_G2_Mix} as a gradient between robustness values of $\bar{\mathcal{G}}_1$ and $\bar{\mathcal{G}}_2$. 
This reveals an important design trade-off, i.e., for a specific diameter requirement, we can utilize the design $\bar{\mathcal{G}}_3$ while affording more robustness than $\bar{\mathcal{G}}_1$ for the same $N$ and $N_L$.

\section{Distributed construction using graph grammars}
\label{subsec:GGs}

This section provides a distributed way of constructing the proposed graphs using graph grammars. Graph grammars are a set of local rules determining interactions between nodes to eventually achieve the desired graph~\cite{klavins2007programmable,yim2007modular,abbas2011hierarchical}. In this method of construction, we provide each node with a label and define a set of rules that dictate how the nodes interact with each other. Moreover, the rules describe how a subset of nodes can create or remove edges amongst them and update their labels accordingly. 
We define a set of rules $\mathcal{R} =\{r_0, r_1, \dots, r_n\}$ through which nodes modify connections with other nodes and update their labels. A rule $r_i$ of the form $G_k$ $\rightharpoonup$  $G_{k+1}$ is applicable to a subgraph $G_k$ representing the state of the subsystem at time step $k$. After applying appropriate rules, the new subgraph $G_{k+1}$ is obtained from $G_k$. 
The graph grammars for $\bar{\mathcal{G}}_1$ and $\bar{\mathcal{G}}_2$ are denoted as $\mathcal{R}_1$ and $\mathcal{R}_2$, respectively. We note that all the nodes are initially labeled $\alpha$ except a seed node, which is labeled $S_1$.
Here, we classify graph grammars into two parts,
\begin{itemize}
    \item $\Pi_1 \xrightarrow{}$ Rules that create edges required for zero forcing and also creates complete graph between leaders.
    \item $\Pi_2 \xrightarrow{}$ Rules that maximize the edge set without compromising SSC.
\end{itemize}
It is important to note that even though the grammars are split into two parts, they can function concurrently with each other. Application of a specific rule only depends on the availability of nodes suitably labelled to apply the rule. Next, we present the graph grammars and also demonstrate their application through examples.

\subsection{Graph Grammars 1 ($\mathcal{R}_1$)}
In this subsection we provide the distributed rules 
$\mathcal{R}_1$ that create graph $\bar{\mathcal{G}}_1$ (as discussed in Section~\ref{subsec:CGG1}) for given number of nodes $N$ and leaders $N_L$.

$\text{$\Pi_1$ :}
\\(r_0)\quad S_i\quad\alpha\:\:\rightharpoonup\:\:L_i\:\:\rule[3pt]{15pt}{1pt}\:\:S_{i+1}\hfill 1\:\leq\:i < N_L\\
(r_1)\quad S_i\:\rightharpoonup\: L_i\hfill i\:=\:N_L\\
(r_2)\quad L_i\quad\alpha\:\:\rightharpoonup\:\:L_i\:\:\rule[2pt]{15pt}{1pt}\:\:\gamma_{i,1}\hfill\forall\:i\\
(r_3)\quad\gamma_{i,j}\quad\alpha\:\:\rightharpoonup\:\:\beta_{i,j}\:\:\rule[2pt]{15pt}{1pt}\:\:\gamma_{i,j+1}\hfill\forall\;\:i,\:1 \leq j < D-1\\
(r_4)\quad \gamma_{i,j}\:\:\rightharpoonup\:\:\beta_{i,j}\hfill j = D-1\\
(r_5)\quad L_i\quad L_j\:\:\rightharpoonup\:\:L_i\:\:\rule[2pt]{15pt}{1pt}\:\:L_j\hfill\forall\:i,j$

$\text{$\Pi_2$ :}
\\(r_6)\quad L_i\quad \beta_{j,1}\:\:\rightharpoonup\:\:L_i\:\:\rule[2pt]{15pt}{1pt}\:\:\beta_{j,1}\hfill \forall\:j\leq\:i\\
(r_7)\quad\beta_{i,j}\quad\beta_{k,j+1}    \:\:\rightharpoonup\:\:\beta_{i,j}\:\:\rule[2pt]{15pt}{1pt}\:\:\beta_{k,j+1}   \hfill \forall j,\:\forall\:k<i\\
(r_8)\quad\beta_{i,j}\quad\beta_{k,j}    \:\:\rightharpoonup\:\:\beta_{i,j}\:\:\rule[2pt]{15pt}{1pt}\:\:\beta_{k,j}\hfill\forall\:i,k\:\:\text{and for each }j$

\vspace{0.1in}
$\Pi_1$ creates $N_L$ number of path graphs, each of size $D$ with the first node being a leader. It also creates a complete graph between all the leader nodes, as in $\mathcal{G}_1$. $\Pi_2$ maximally adds edges to the graph while ensuring that the graph remains strong structurally controllable through the leader nodes. 
Figure~\ref{fig:GG}(a) illustrates an example of the execution of $\mathcal{R}_1$ to generate $\bar{\mathcal{G}}_1$ for $N_L = 3$ and $N = 12$.

\subsection{Graph Grammars 2 ($\mathcal{R}_2$)}
Here, we provide the set of distributed rules $\mathcal{R}_2$ that create graph $\bar{\mathcal{G}}_2$ (as discussed in Section~\ref{subsec:CGG2}) for given number of nodes $N$ and leaders $N_L$.

$\text{$\Pi_1$ :}
\\(r_0)\quad S_i\quad \alpha\:\:\rightharpoonup\:\:L_i\:\:\rule[3pt]{15pt}{1pt}\:\:S_{i+1}\hfill 1\:\leq\:i\:< N_L\\
(r_1)\quad S_i\:\rightharpoonup\: L_i\hfill i\:=\:N_L\\
(r_2)\quad L_1\quad \alpha\:\:\rightharpoonup\:\:L_1\:\:\rule[3pt]{15pt}{1pt}\:\:\beta_1\\
(r_3)\quad \beta_i\quad \alpha\:\:\rightharpoonup\:\:\gamma_i\:\:\rule[3pt]{15pt}{1pt}\:\:\beta_{i+1}\hfill 1\:\leq\:i\:<\:(N-N_L)\\
(r_4)\quad \beta_i\:\:\rightharpoonup\:\:\gamma_i\hfill i\:=\:(N-N_L)\\
(r_5)\quad L_i\quad L_j\:\:\rightharpoonup\:\:L_i\:\:\rule[2pt]{15pt}{1pt}\:\:L_j\hfill\forall\:i,j$

$\text{$\Pi_2$ :}
\\(r_6)\quad L_i\quad\gamma_i\:\:\rightharpoonup\:\:L_i\:\:\rule[3pt]{15pt}{1pt}\:\:\gamma_i\hfill \forall \:i\neq 1\;
$

\begin{figure*}[htb]
    \centering
    \begin{subfigure}[b]{0.75\textwidth}
	 \centering
    \includegraphics[scale=0.19]{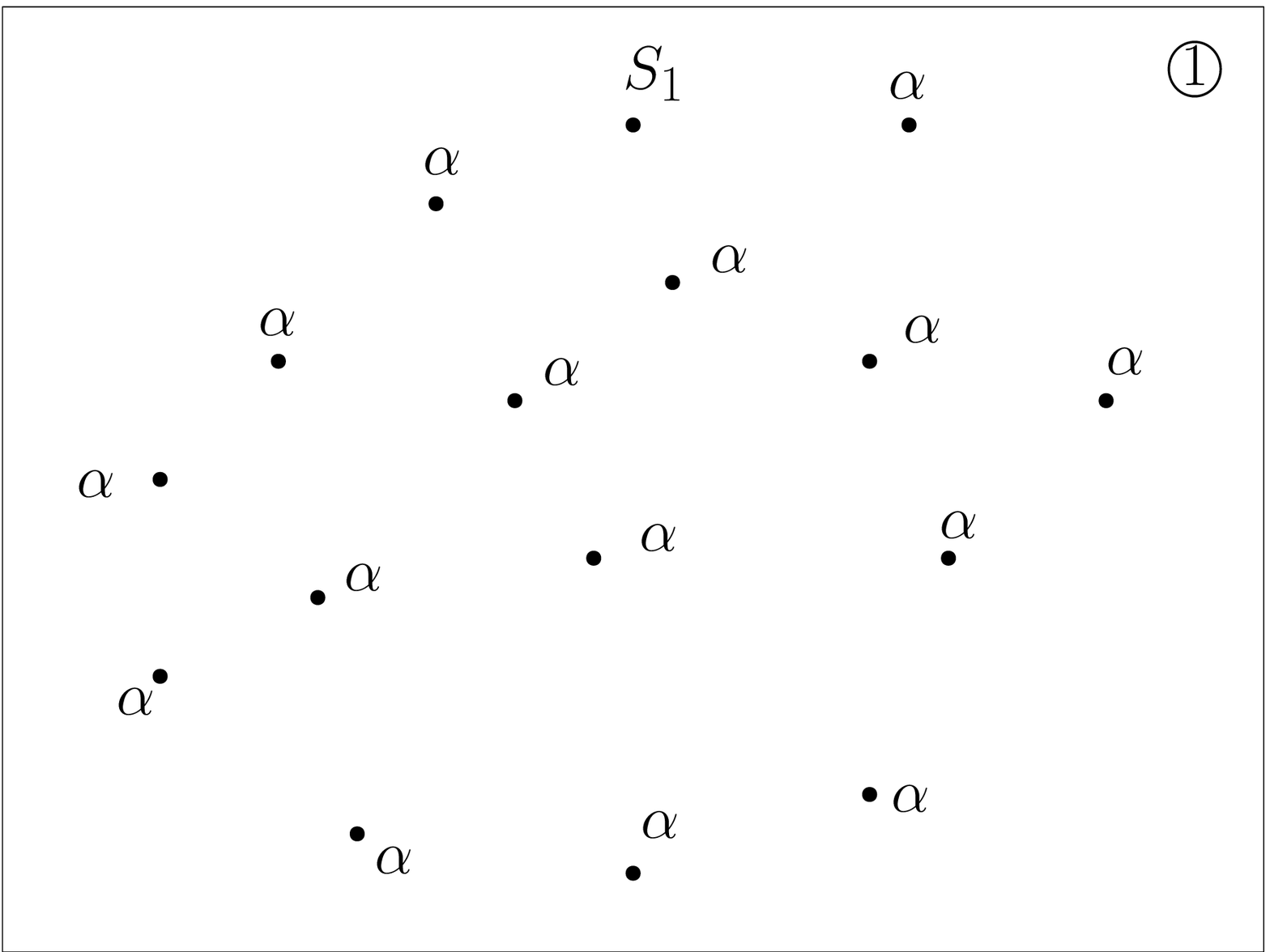}\;%
    \includegraphics[scale=0.19]{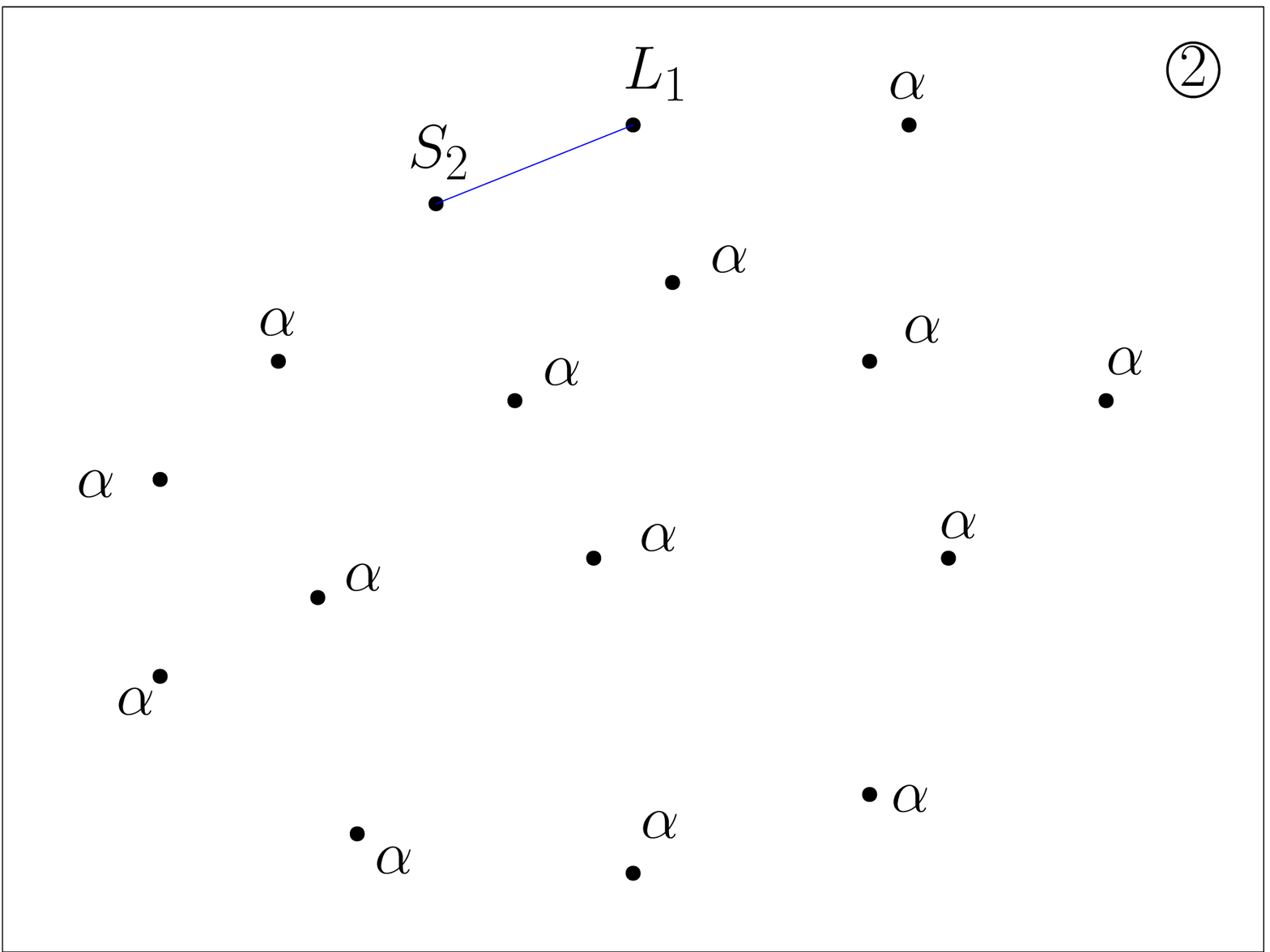}\;%
    \includegraphics[scale=0.19]{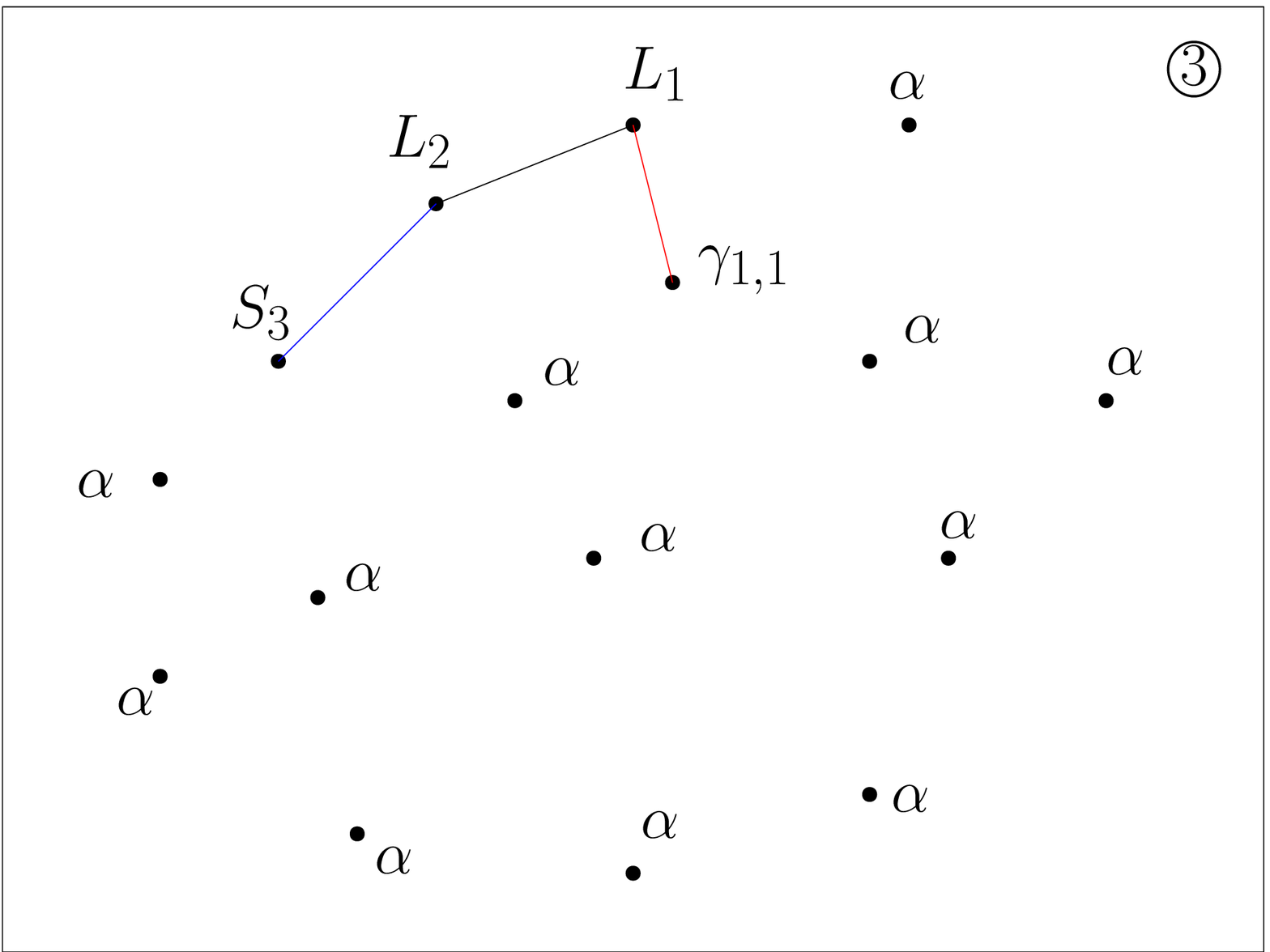}\;%
   \includegraphics[scale=0.19]{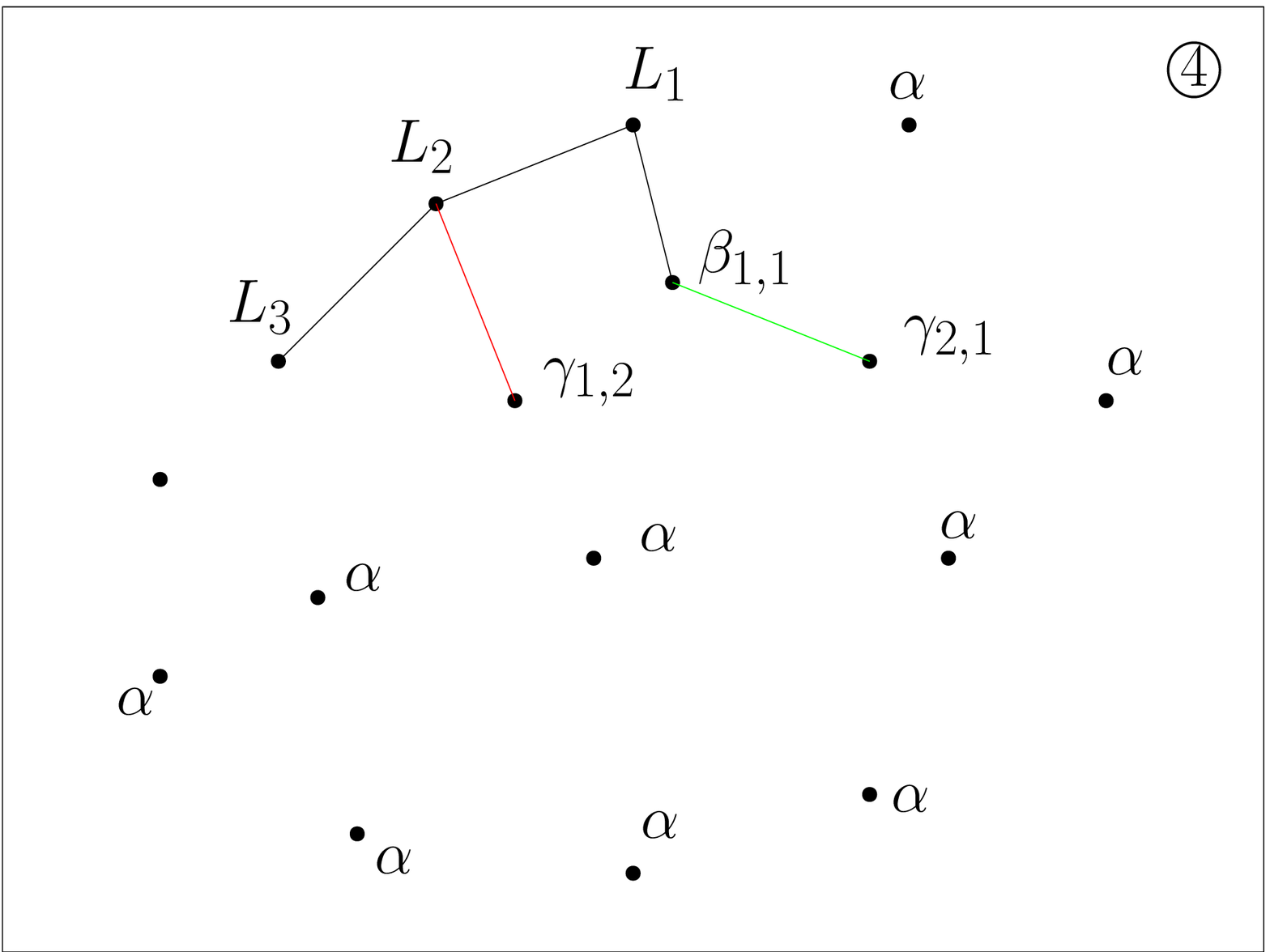}\\%
    \includegraphics[scale=0.19]{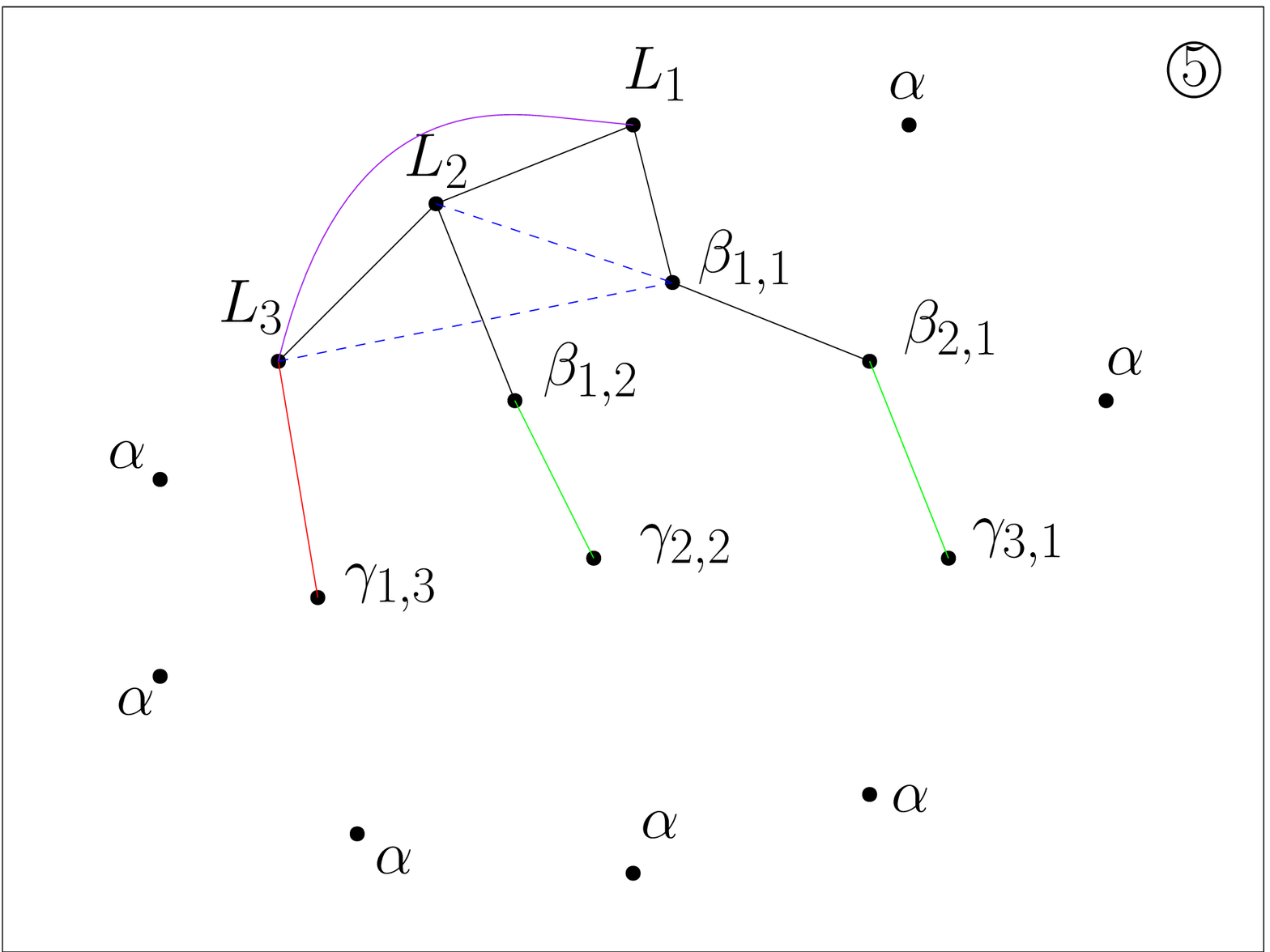}\;%
   \includegraphics[scale=0.19]{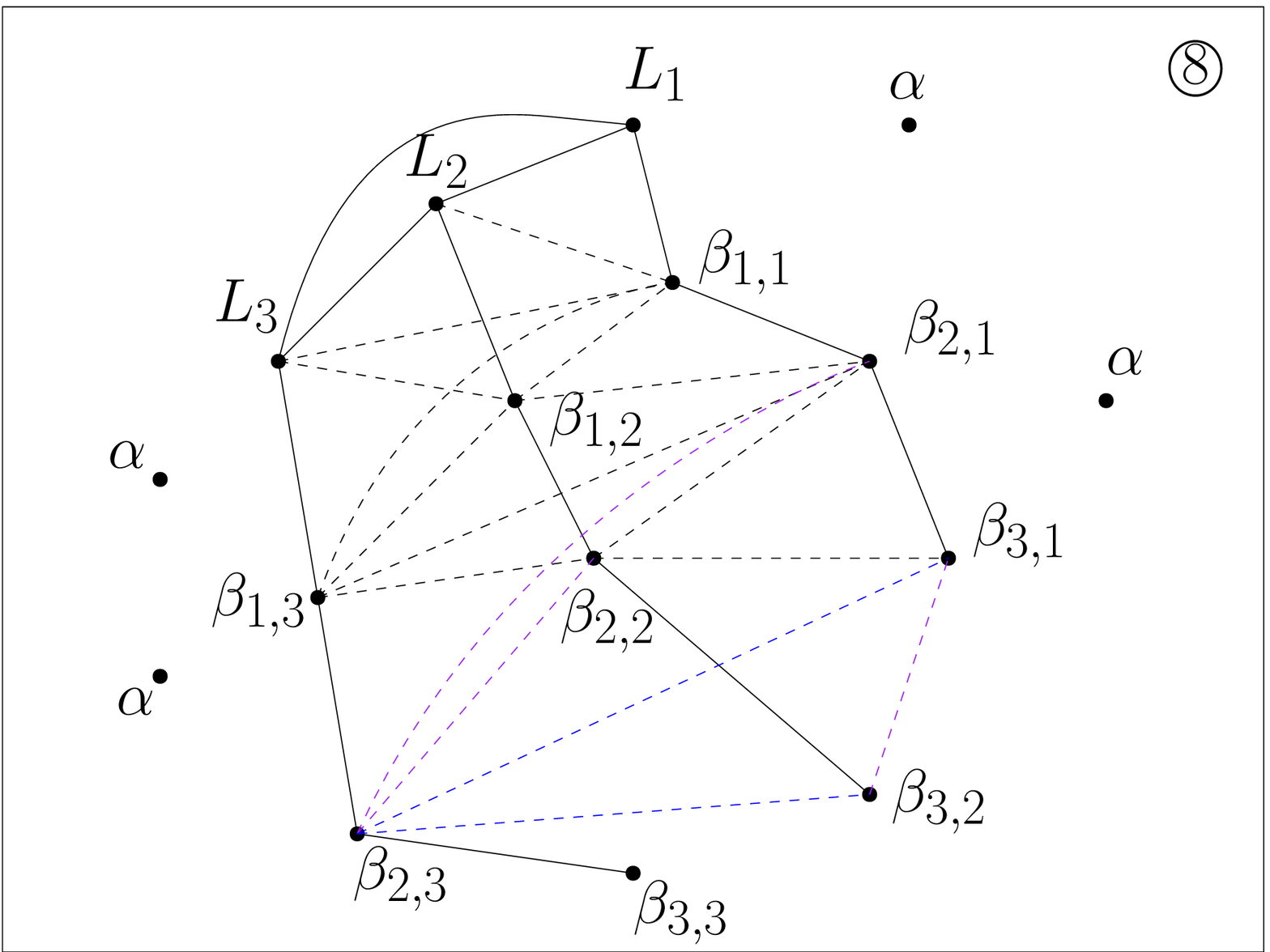}\;%
    \includegraphics[scale=0.19]{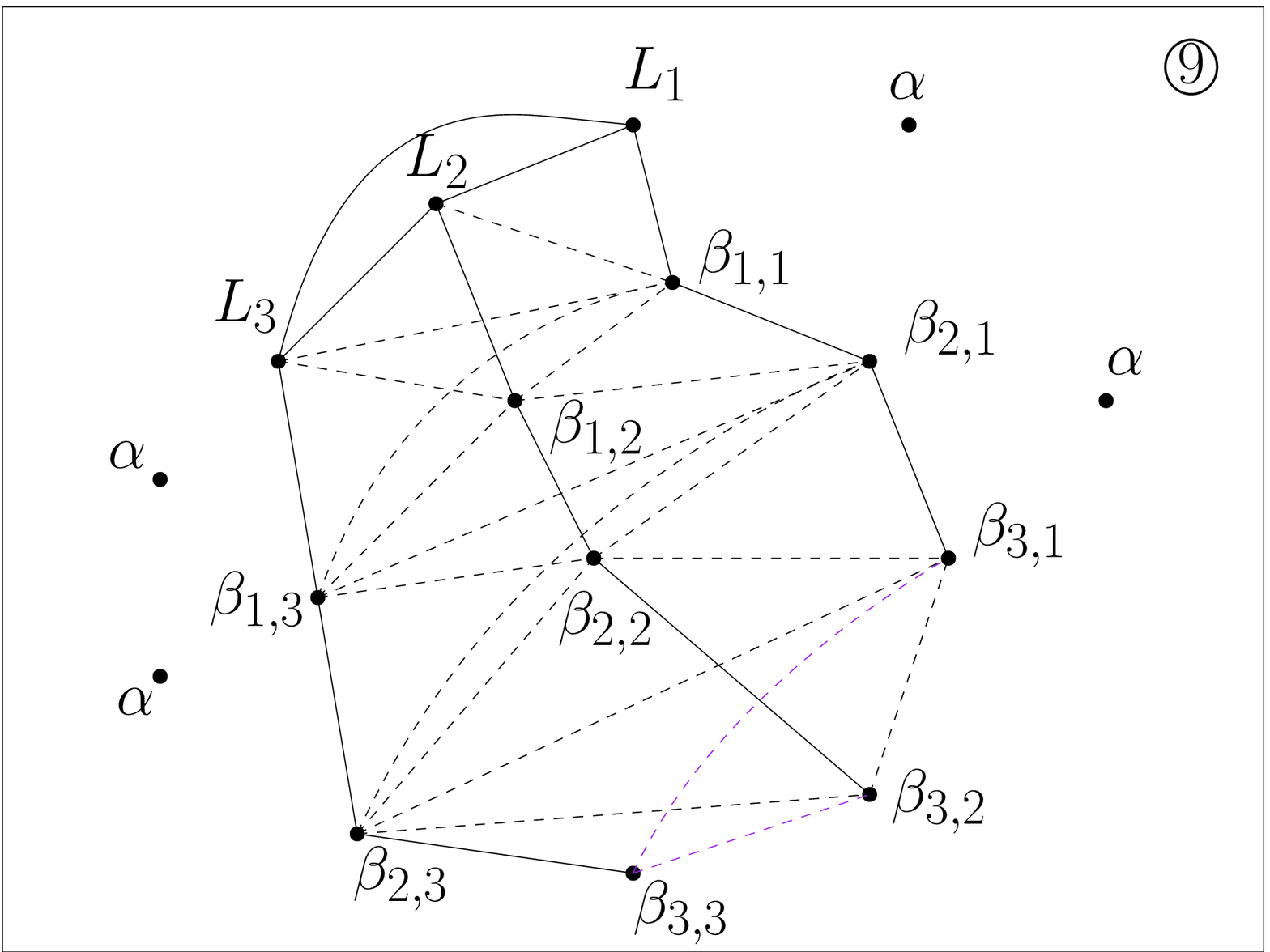}\;%
   \includegraphics[scale=0.19]{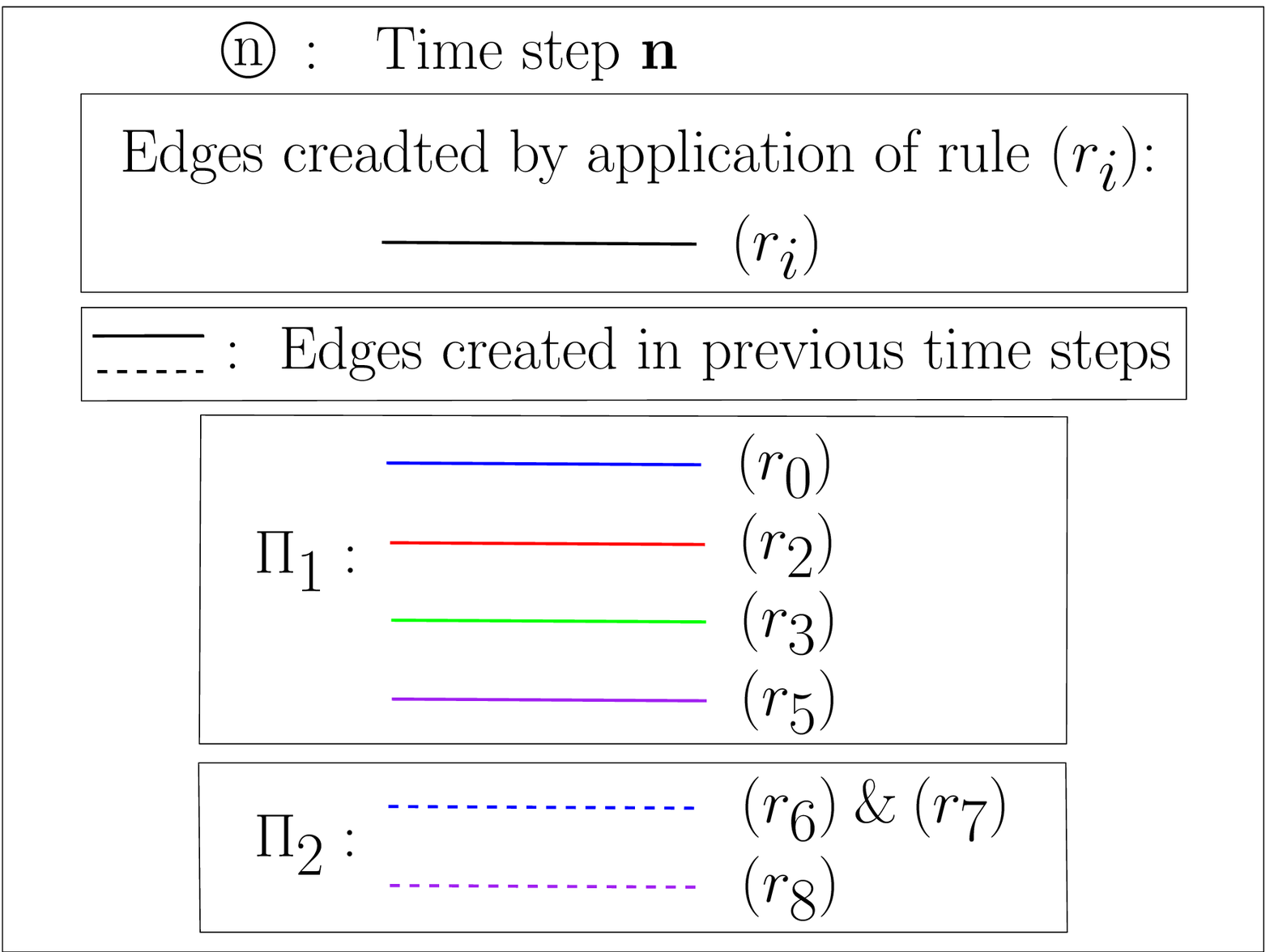}%
    \caption{$\bar{\mathcal{G}}_1$}
	\end{subfigure}\\
	\begin{subfigure}[b]{0.75\textwidth}
\centering
  \includegraphics[scale=0.19]{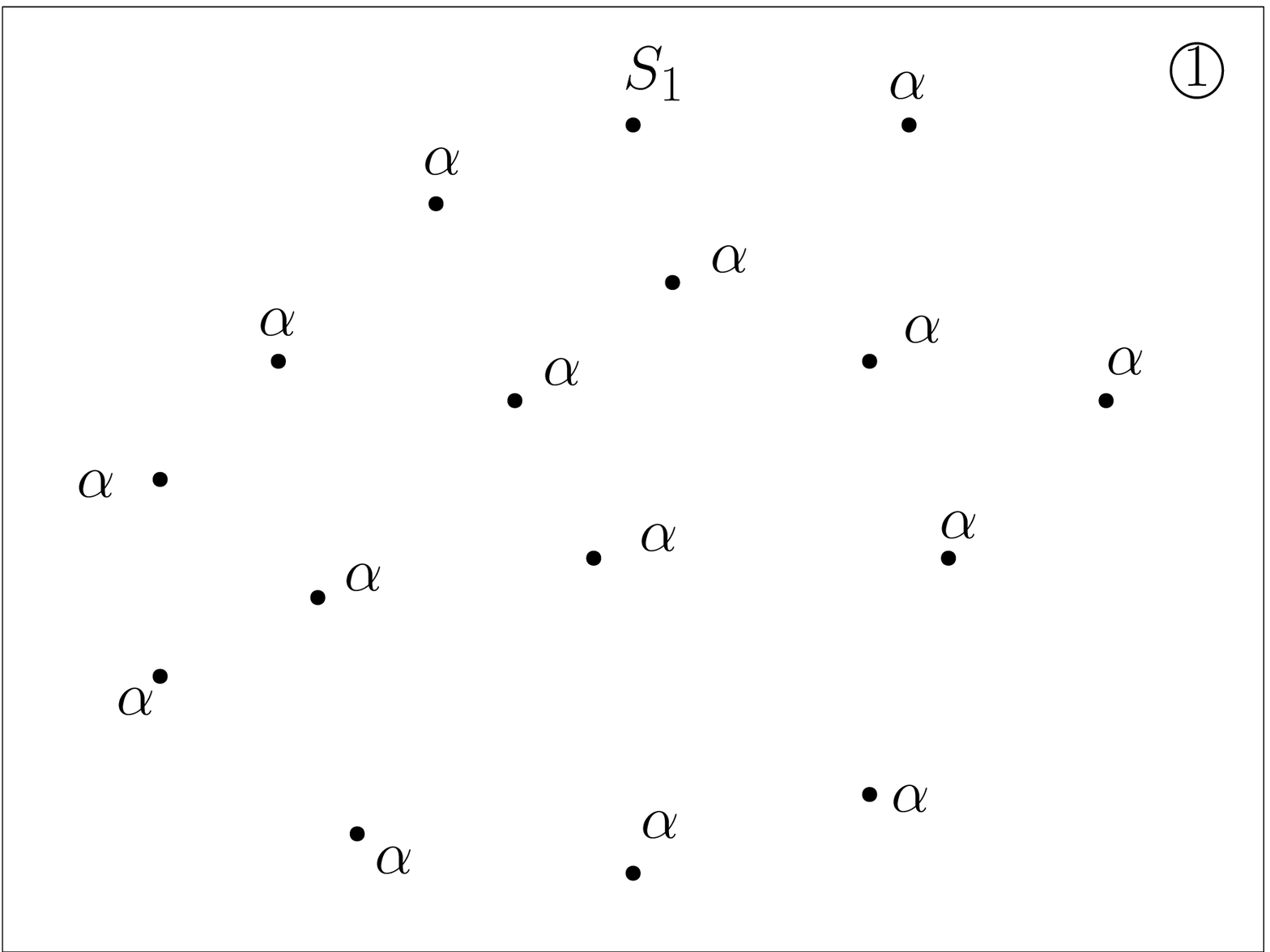}\;%
    \includegraphics[scale=0.19]{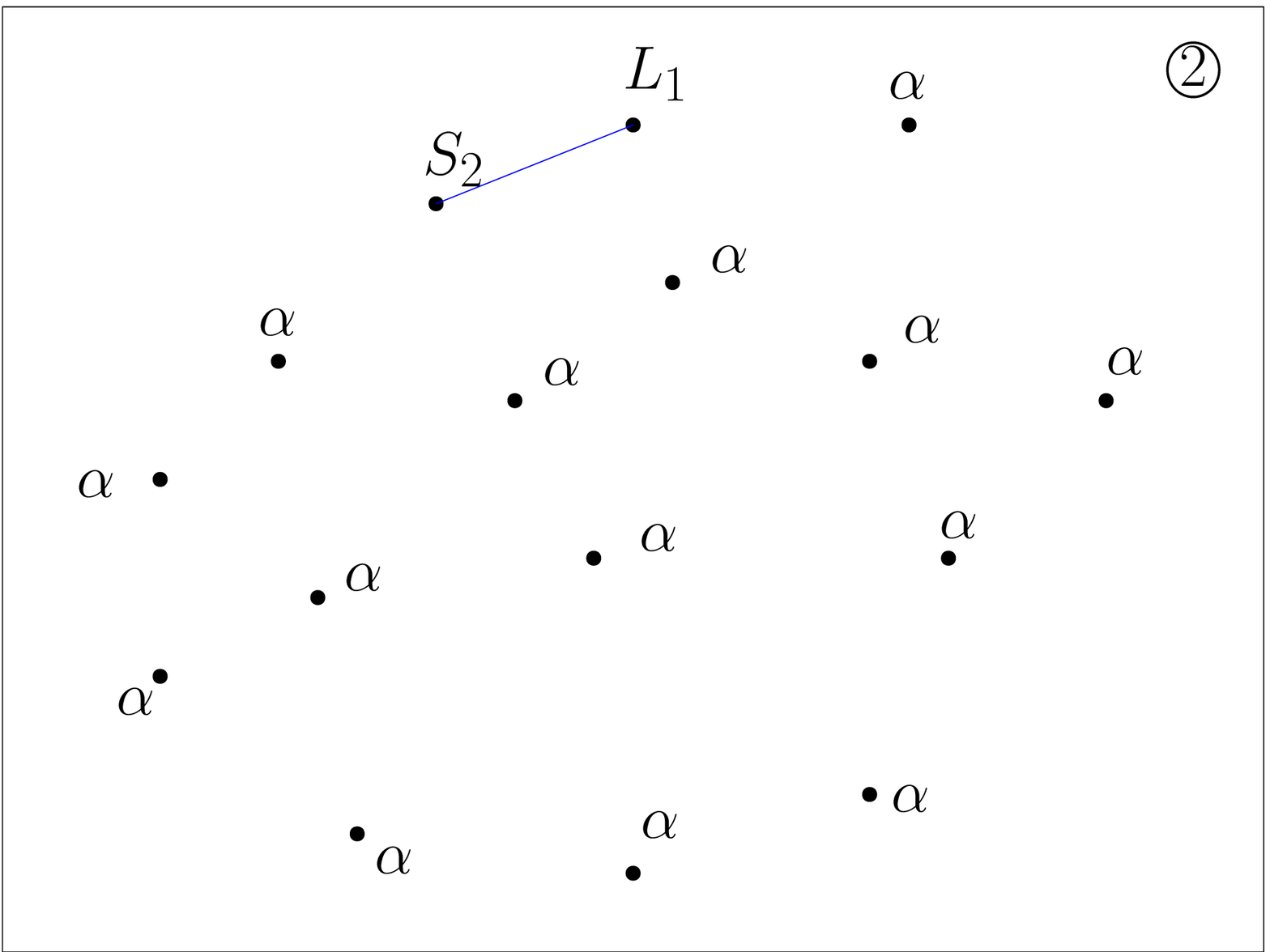}\;%
    \includegraphics[scale=0.19]{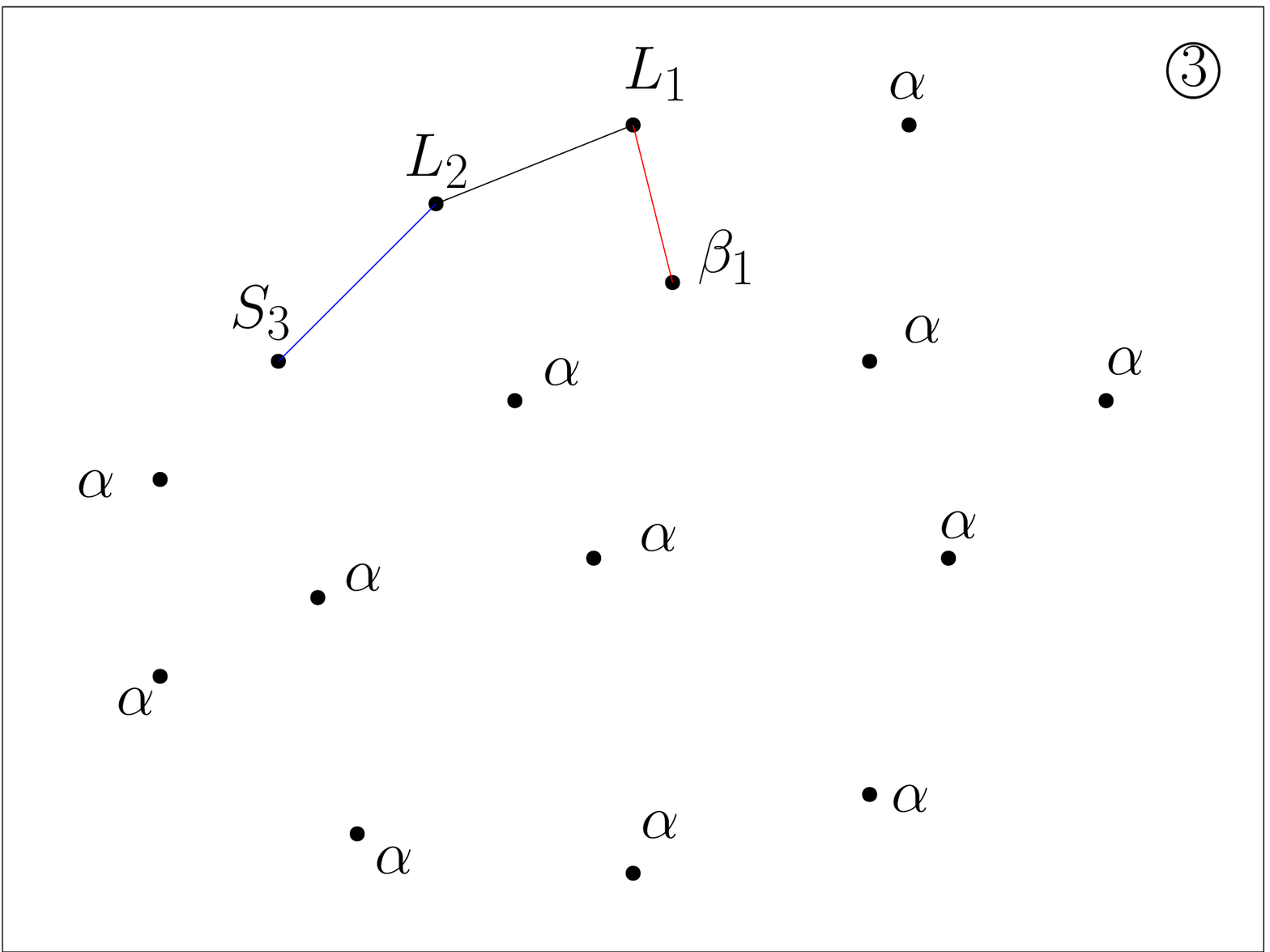}\;%
   \includegraphics[scale=0.19]{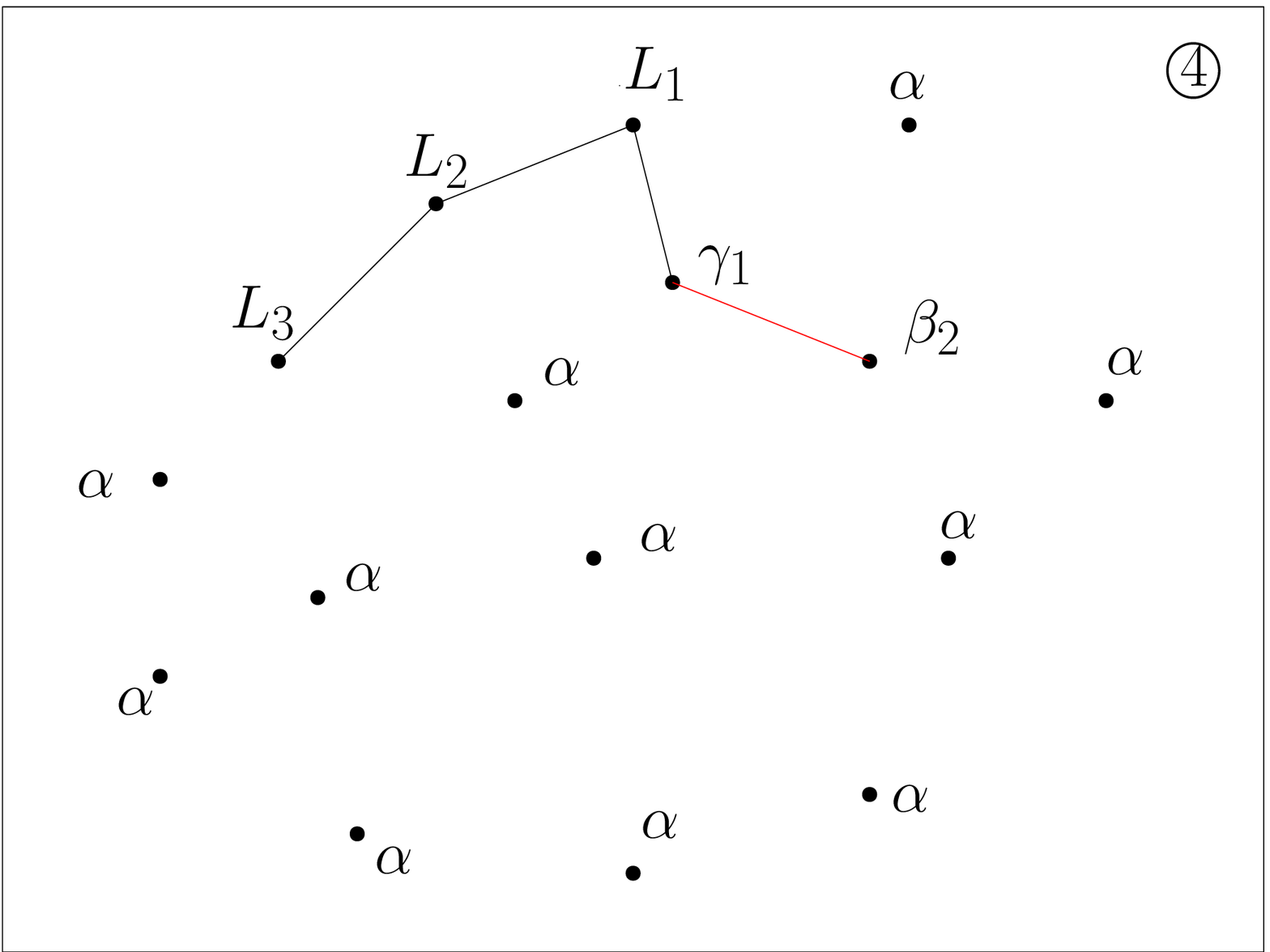}\\%
  \includegraphics[scale=0.19]{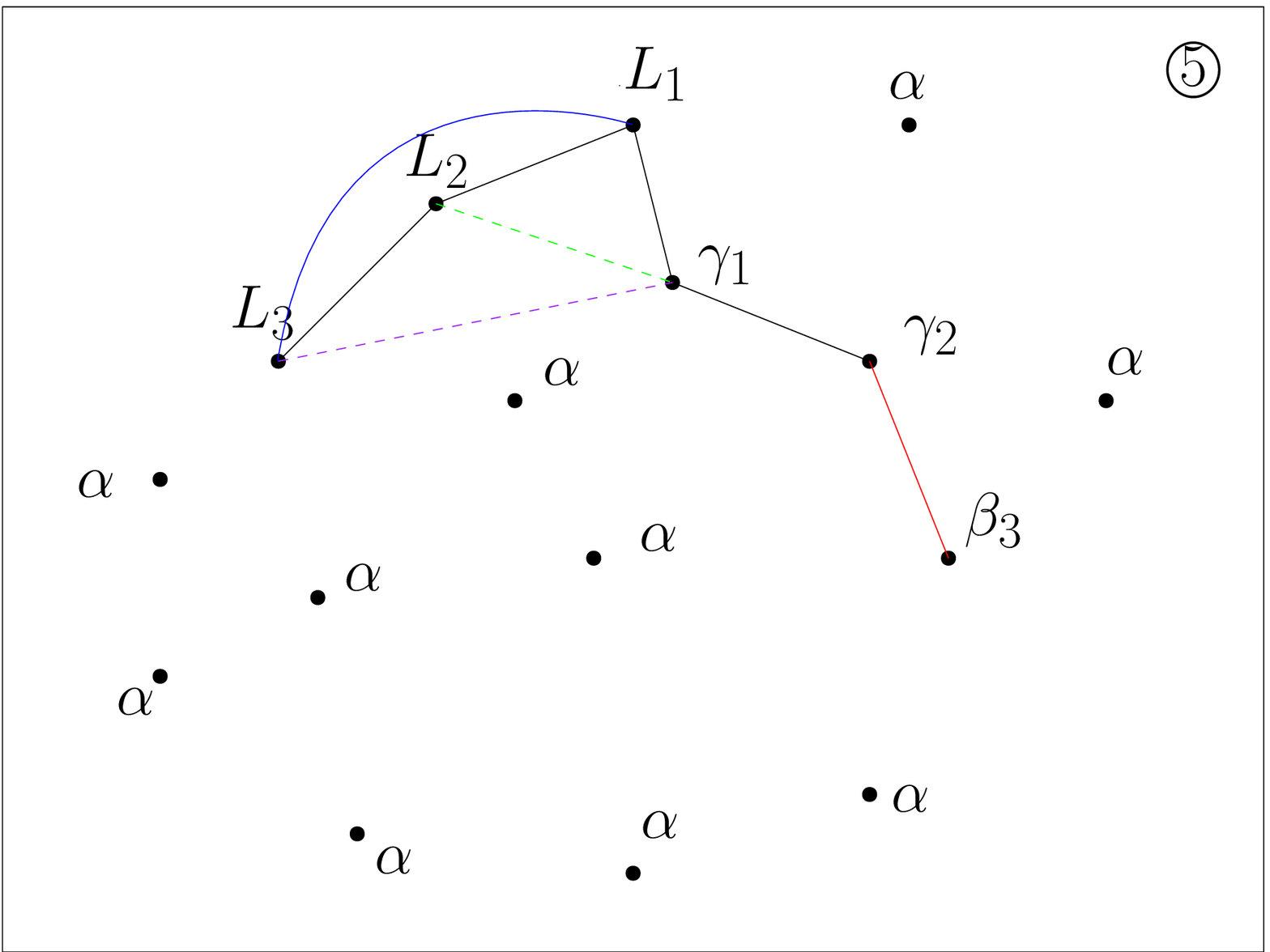}\;%
   \includegraphics[scale=0.19]{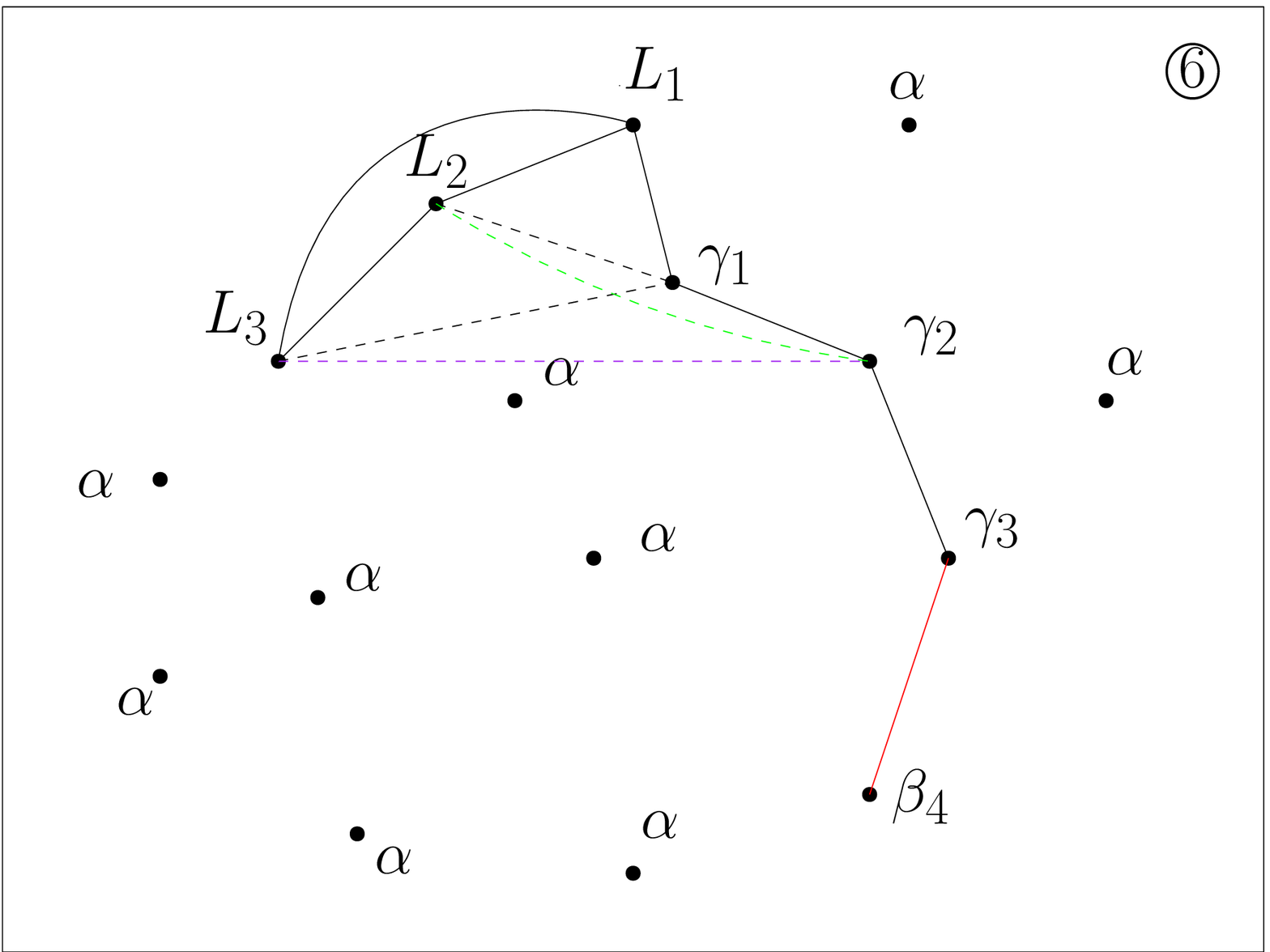}\;%
    \includegraphics[scale=0.19]{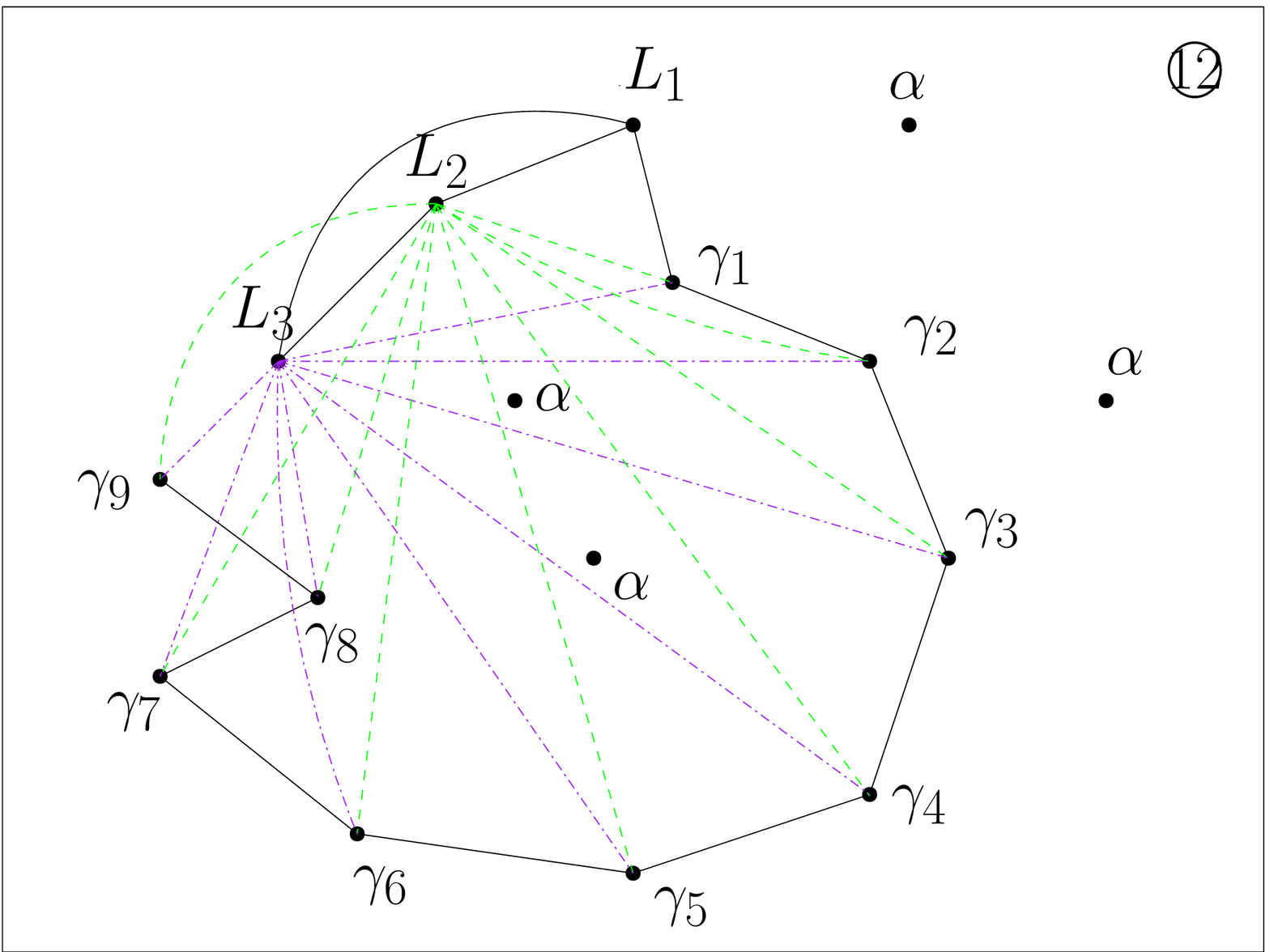}\;%
    \includegraphics[scale=0.19]{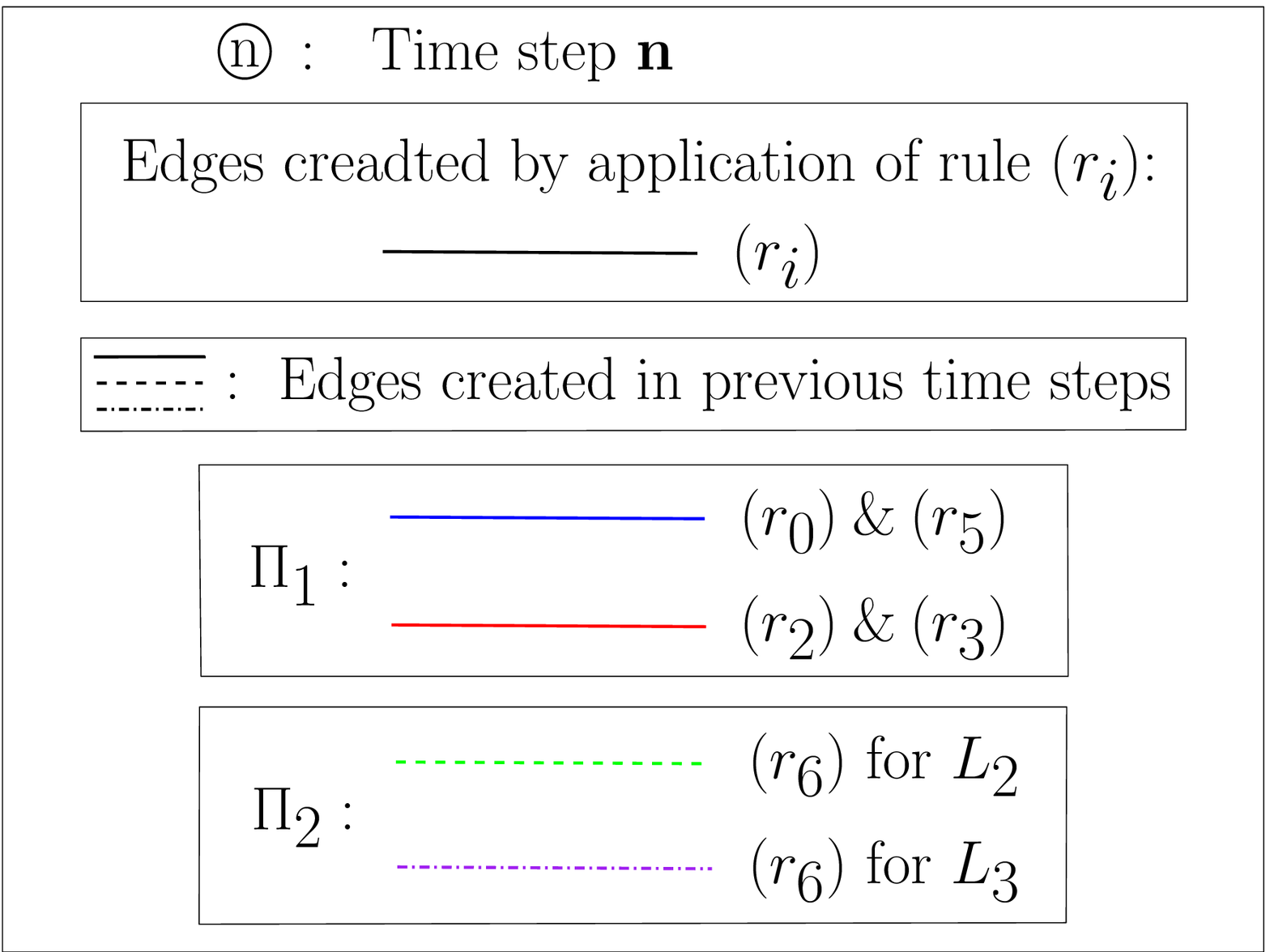}%
    \caption{$\bar{\mathcal{G}}_2$}
	\end{subfigure}
    \caption{(a) Construction of $\bar{\mathcal{G}_1}$ using $\mathcal{R}_1$. (b) Construction of $\bar{\mathcal{G}_2}$ using $\mathcal{R}_2$.}
    \label{fig:GG}
\end{figure*}

$\Pi_1$ creates a path graph of size $N - N_L + 1$, where the first node is the first leader. It also creates a complete graph between all the leader nodes. On the other hand, $\Pi_2$ adds maximal edges to the graph without affecting the SSC. 
An example of $\bar{\mathcal{G}_2}$ with $N_L = 3$ and $N = 12$ constructed using $\mathcal{R}_2$ is shown in Figure~\ref{fig:GG}(b).



\section{Conclusion}
\label{sec:conclusion}

In this work, we presented multiple ways of constructing graphs that are strong structurally controllable, and maximally robust. Using the notion of zero forcing, we exploited the relation between controllability and robustness. 
We provided three different network designs for the given number of nodes $N$, the number of leaders $N_L$, and the diameter $D$. These designs exhibited different properties in terms of robustness and diameter; however, they all were strong structurally controllable with given parameters. For arbitrary $N$ and $N_L$, our proposed graphs $\bar{\mathcal{G}}_1$ and $\bar{\mathcal{G}}_2$ were of diameters $N/N_L$ and $2$, respectively. On the other hand, the network design $\bar{\mathcal{G}}_3$ enabled selecting any diameter value between $2$ and $N/N_L$, thus, generating graphs with different robustness. 
Moreover, we provided distributed ways of constructing these graphs using graph grammars. These local rules determined the interactions between nodes, and created desired graphs $\bar{\mathcal{G}}_1$ and $\bar{\mathcal{G}}_2$ in a decentralized manner.

\bibliographystyle{IEEEtran} 
\bibliography{references}  
\end{document}